\documentclass[journal,10pt,twocolumn]{IEEEtran}

\usepackage[cmex10]{amsmath}
\usepackage{amssymb}
\usepackage{amsthm}
\usepackage{graphicx}
\usepackage{color}
\usepackage{bm}
\usepackage{bbm}
\usepackage{url}
\usepackage{array}
\usepackage{multicol}
\usepackage{algorithm}
\usepackage[colorlinks]{hyperref}
\usepackage{url}
\usepackage{algpseudocode}

\theoremstyle{remark}
\newtheorem{remark}{Remark}
\usepackage{epstopdf}

\usepackage{array}
\makeatletter
\newcommand{\thickhline}{%
    \noalign {\ifnum 0=`}\fi \hrule height 1pt
    \futurelet \reserved@a \@xhline
}
\newcolumntype{"}{@{\hskip\tabcolsep\vrule width 1pt\hskip\tabcolsep}}
\makeatother

\begin{document}

\title{Fast Fluid Antenna Multiple Access}

\author{\IEEEauthorblockN{Noor Waqar, %\emph{Student Member, IEEE}, 
		                           Kai-Kit Wong, \emph{Fellow, IEEE},
		                           Chan-Byoung Chae, \emph{Fellow, IEEE}, and 
		                           Ross Murch, \emph{Fellow, IEEE}
}
\vspace{-8mm}

%\thanks{For the purpose of open access, the authors will apply a Creative Commons Attribution (CC BY) licence to any Author Accepted Manuscript version arising.}

\thanks{The work of K. K. Wong is supported by the Engineering and Physical Sciences Research Council (EPSRC) under grant EP/W026813/1.}
\thanks{The work of C. B. Chae is supported by the Institute for Information and Communication Technology Planning and Evaluation (IITP)/National Research Foundation of Korea (NRF) grant funded by the Ministry of Science and ICT (MSIT), South Korea, under Grant RS-2024-00428780 and 2022R1A5A1027646.}
\thanks{The work of R. Murch was supported by the Hong Kong Research Grants Council Area of Excellence Grant AoE/E-601/22-R.}

\thanks{N. Waqar and K. K. Wong are with the Department of Electronic and Electrical Engineering, University College London, London WC1E 7JE, United Kingdom. K. K. Wong is also affiliated with Yonsei Frontier Laboratory, Yonsei University, Seoul, 03722, Republic of Korea.}
\thanks{C. B. Chae is with School of Integrated Technology, Yonsei University, Seoul, 03722, Republic of Korea.}
\thanks{R. Murch is with the Department of Electronic and Computer Engineering, Hong Kong University of Science and Technology, Clear Water Bay, Hong Kong SAR, China.}

\thanks{Corresponding author: Kai-Kit Wong (e-mail: $\rm kai\text{-}kit.wong@ucl.ac.uk$).}
}
\maketitle

\begin{abstract}
Fast fluid antenna multiple access (FAMA) is an idea that promises to overcome severe interference in massive access scenarios by reconfiguring the antenna's position at the receiver side on a symbol-by-symbol basis, without the need of precoding nor any other interference mitigation techniques. However, this idea is commonly studied under a \emph{genie-aided} premise: each user terminal (UT) can probe \emph{all} fluid-antenna ports in every symbol instance and ideally knows the instantaneous signal-interference split for the received signals at all the ports. Such assumption is unrealistic since it implies impractical hardware and switching limits, pilot overhead, as well as an unknown ability to determine the signal-interference split. This paper revisits the fast FAMA communication problem and asks a key question: can a UT act \emph{as if} it had full per-port interference knowledge while observing only a small fraction of ports? To this end, we propose a \emph{copula-aided FAMA} framework that learns the joint dependence structure of the complex triplets $(r_k,h_k,I_k)$ across ports, where $r_k$, $h_k$ and $I_k$ denote, respectively, the received signal, the channel coefficient and the aggregate interference signal at the $k$-th port, and uses this learned model to infer unobserved channels and interference. Concretely, we devise an attention-copula time-series model that is trained under random partial-observation masks and evaluated under both rich and finite-scattering channel models. Simulation results indicate that the reconstruction normalized mean-square-error (NMSE) for  $h$, $r$, and $I$ drops to the order of $10^{-4}$ once the number of observed ports, $M$, exceeds the spatial degrees of freedom (DoF). We further provide the binary symmetric channel (BSC) system sum-rate, illustrating that the proposed approach matches an oracle benchmark with perfect per-port interference knowledge. These results suggest that near-ideal fast FAMA be attainable with modest per-symbol sensing, substantially relaxing the need for genie-aided interference observability.
\end{abstract}

\begin{IEEEkeywords}
Fluid antenna system (FAS), fast fluid antenna multiple access (FAMA), massive connectivity.
\end{IEEEkeywords}

\vspace{-2mm}
\section{Introduction}
\subsection{Background}
\IEEEPARstart{T}{he explosive} growth of connected devices and data-hungry applications such as holographic media, extended reality, and the Internet of Everything (IoE) is pushing wireless systems toward unprecedented levels of spectrum reuse and connection density \cite{quy2025strategic,wang2023ontheroad,jiang2021theroad,chowdhury20206Gwireless}. Future sixth-generation (6G) wireless networks are required to deliver hyper-reliable, low-latency connectivity to a large number of user terminals (UTs), over a highly congested wireless medium, sometimes even without infrastructure such as base station (BS). Achieving this hinges critically on how aggressively we are able to reuse the spectrum while mitigating interference \cite{clerckx2024multiple,liu2024theroad}.

Apparently, multiuser multiple-input multiple-output (MU-MIMO) and its extra-large version \cite{wang2024XLMIMO} will continue to be the technology to lift the spectral efficiency and support massive access. Nevertheless, the increasing overhead for channel state information (CSI) acquisition and feedback plus the complexity for large-scale precoding optimization will be the obstacle going forward \cite{Villalonga2022spectral}. More worryingly, the high peak-to-average power ratio (PAPR) resulting from precoding greatly increases power consumption of the BS by reducing the efficiency of power amplifiers (PAs) \cite{Hung-2014}. Recent 6G discussion thus seems to favor having only $32$ antennas at each BS, a step down from $64$ antennas in the fifth-generation (5G) systems.

Often labelled as next-generation multiple access (NGMA) schemes, non-orthogonal multiple access (NOMA) \cite{ahmed2024unveil,ding2024design}, and rate-splitting multiple access (RSMA) \cite{mao2022rsma,Clerckx2023aprimer}, have in recent years emerged as a massive connectivity solution due to its aggressive reuse of spectrum. Like MU-MIMO precoding, their success depends on accurate CSI at the transmitter, and worse successive interference cancellation (SIC) at the receiver ends, which raises doubts on their practicality.

\vspace{-2mm}
\subsection{Scalable Multiple Access}
A truly scalable multiple access scheme is long overdue---one that can serve many devices or UTs on the same channel-use but with simple signal processing. Preferably,
\begin{itemize}
\item There should only be simple or no precoding at the BS, which reduces the burden of CSI acquisition and keeps the PAPR low to ensure power-efficient MIMO systems. If precoding is not used, then a centralized infrastructure such as BS is not required. Such scheme will be ideal for device-to-device (D2D) communications, or massive IoE scenarios where centralized optimization is infeasible.
\item There should be no SIC at the receiver ends because SIC can never be reliably performed in practice if a massive number of co-channel UTs are present.
\item Power control over the UTs is also not welcome because the optimization will be prohibitively complex. 
\end{itemize}

Satisfying the above conditions certainly results in a scalable multiple access technology but then {\em how interference can be addressed?} The emerging fluid antenna system (FAS) concept offers an idea to tackle interference under these situations.

First introduced in \cite{wong2020performance,wong2021fluid}, FAS is a system concept that treats the antenna as a reconfigurable physical-layer resource unit to broaden system design and network optimization \cite{new2024tutorial,hong2025contemporary,New-2026jsac}. In \cite{Lu-2025}, Lu {\em et al.}~provided an explanation of FAS through the lens of electromagnetics. FAS is hardware agnostic and can appear in many forms, from movable elements \cite{zhu2024historical}, to liquid-based antennas \cite{shen2024design,Shamim-2025}, metamaterials \cite{Liu-2025arxiv,Zhang-jsac2026}, reconfigurable pixel-based designs \cite{zhang2024pixel,tong-2025pixel,Wong-wc2026}. Focusing on position reconfigurability, tremendous diversity in FAS has been reported in \cite{new2024insights,ramirez2024new,new2023information,zhu2025geometric}. %Copula-based analysis has been proven effective in handling the channel correlation between the FAS ports \cite{ghadi2023copula,ghadi2023gaussian,tlebaldiyeva2023copula,r2024gaussian}.

Beyond diversity benefits, FAS has introduced new ways for multiple access, broadly referred to as fluid antenna multiple access (FAMA) \cite{Shah2024survey}. The concept of FAMA was first proposed by Wong {\em et al.}~in \cite{wong2022fama} which argues that interference can be avoided entirely by utilizing the spatial opportunity where the aggregate interference signal suffers from a deep fade naturally resulting from the fading phenomenon. Technically, at the $k$-th port of FAS, the UT of interest has the received signal
\begin{equation}\label{eq:rk0}
r_k=h_ks_u+\underbrace{\sum_{\bar{u}\ne u}h_k^{(\bar{u})}s_{\bar{u}}}_{I_k}+\eta_k,
\end{equation}
in which $s_u$ is the information-bearing symbol to be retrieved from the received signal, $s_{\bar{u}}$ denotes the symbols from the interferers for $\bar{u}\ne u$, $h_k$ represents the channel coefficient between the transmitter and the $k$-th port of the UT, $h_k^{(\bar{u})}$ is the channel from interferer $\bar{u}$ to the $k$-th port of the UT of interest, and $\eta_k$ denotes the additive noise.

The received signals at the FAS ports, $r_1,r_2,\dots,r_K$, are correlated but the interference terms, $I_1,I_2,\dots,I_K$, fluctuate because of how the interfering signals combine over different positions. Importantly, if the number of interferers is large, $I_k$ behaves like a complex Gaussian noise and $|I_k|$ is Rayleigh distributed. As a result, deep fade of $I_k$ exists meaning that there can be one or more ports where $I_k\approx 0$. Encouraged by this observation, \cite{wong2022fama} proposed to switch to the optimal port, $k^{\star}$, for receiving the signal such that
\begin{equation}\label{eq:ffama-select}
k^\star=\arg\max_k\frac{|h_k|^2}{|I_k|^2},
\end{equation}
where the noise term is dropped for simplicity. At port $k^\star$, the aggregate interference naturally `disappears'.

Since $|I_k|$ is data-dependent (a function of the transmitted symbols by the interferers), the optimal port $k^\star$ changes on a symbol-by-symbol basis. This approach was later referred to as {\em fast} FAMA in \cite{wong2023fFAMA}. The fast FAMA scheme \cite{wong2022fama} described above can mitigate interference from many interfering sources without any of power control, precoding and SIC \cite{Wang-wc2024}. Despite this, fast FAMA is deemed unrealistic because
\begin{itemize}
\item It requires to have access to the received signals from all $K$ FAS ports on every symbol instance. This means that the FAS needs to switch $K$ times faster than the symbol rate on a single radio-frequency (RF) chain to receive the signal samples at all the ports, or the UT receiver needs to have $K$ RF chains to receive all the signals at once.
\item The knowledge of full CSI, $\{h_k\}_{\forall k}$, is needed but estimating it can be expensive when $K$ is large.
\item More critically, it requires the knowledge of $\{I_k\}$ at all the ports so that (\ref{eq:ffama-select}) can be performed. Even when $r_k$ is known, it is not understood how $I_k$ can be estimated.
\end{itemize}

Because of the above concerns, several variants of FAMA have been developed. In \cite{wong2023fFAMA}, techniques for estimating $\{I_k\}$ were proposed but considerable performance degradation was resulted. Subsequently, the work in \cite{wong2023sFAMA,xu2024revisiting,hong2025downlink} proposed the slow FAMA scheme that selects the optimal port based on maximizing the average signal-to-interference plus noise ratio (SINR) which does not depend on the interferers' symbols. In so doing, the UT only changes its receiving port when the CSI changes. However, the price is a much reduced multiplexing capability, and a much small number of UTs can be supported comparing to fast FAMA. To improve this, \cite{wong2024cuma} developed the compact ultra-massive array (CUMA) scheme where instead of selecting one `best' port, it employs a multi-port combining approach in the analog domain \cite{rao2025geometric}, which can improve the slow FAMA scheme considerably. Besides, channel coding has been shown to be synergistic with FAMA \cite{hong2025coded,hong20255gcoded}. In \cite{waqar2025turbocharging}, progress has been made to address the issues of fast FAMA. Specifically, turbo FAMA uses port-shortlisting rules, performs processing in the data domain, adopts diffusion models and deep joint source-channel coding (JSCC), to alleviate the need for knowing $\{I_k\}$, resulting in extraordinary multiplexing gain. Nonetheless, the processing complexity at the UT is high and the assumptions of full received signals $\{r_k\}$ and CSI $\{h_k\}$ remain. To restore the conceptual simplicity of fast FAMA and make (\ref{eq:ffama-select}) practical, we aim to answer the question:

\vspace{1mm}
\begin{quote}
\begin{center}
\emph{Can a FAS-equipped UT behave \textbf{as if} it had access to the interference field over all ports $\{I_k\}$, while actually knowing only a small fraction of the received signals and CSI, $\{r_k\}$ and $\{h_k\}$?}
\end{center}
\end{quote}
\vspace{1mm}

We argue that the answer is positive if one leverages the rich spatial structure within the FAS channels through \emph{probabilistic modeling}. The interference-plus-noise signal observed at different ports and symbols is highly correlated and strongly shaped by the underlying scattering geometry. This structure has already been exploited in analytical performance studies via advanced correlation models and copula-based approaches \cite{ghadi2023copula,ghadi2023gaussian,tlebaldiyeva2023copula,r2024gaussian}. However, these works mostly treat the channel statistics as objects to be analyzed, not as ingredients in a \emph{learned} model that can perform real-time inference.

\vspace{-2mm}
\subsection{Inference for FAS and FAMA Systems}
CSI estimation for FAS has recently been tackled in \cite{new2025fas_channel_est,zhang2020successive_fas_ce,xu2024sbl_fas_ce} and it was found that channel sparsity and correlation structures are important features that can greatly simplify the estimation process. At the same time, there has been a surge of interest in artificial intelligence (AI)-driven designs for FAS and FAMA. Deep neural networks and reinforcement learning have been used to predict good ports, estimate port SINR maps, or perform scheduling in slow FAMA scenarios \cite{waqar2023deep,eskandari2024cgan_slowfama,waqar2024oppor_fama}. AI-enabled frameworks have been proposed more broadly for FAS optimization, including graph neural networks and large language models (LLMs) that operate at the system level \cite{he2025graph,wang2025largex,deng2025llmfassurvey,guo2025llmfas,portllm2025,yang2025fasllm,Wang-2026wcl}. 

However, most existing AI-based schemes treat the port SINR (or related metrics) as \emph{targets} that must be provided or approximated, rather than as \emph{random variables} to be inferred jointly from partial observations. Put differently, they learn discriminative mappings but do not explicitly model the \emph{joint distribution} of the desired channel, interference, and received signal over all ports. As a consequence, they are not naturally equipped to answer questions such as what is the posterior distribution of the instantaneous interference $I_k$ at every port $k$ given noisy observations of $(r,h)$ at only a subset of ports and how confident is the system that the currently chosen port or port set is near-optimal in terms of instantaneous SINR?

In relation to this, previous studies have embraced copula theory to separately model marginal fading behavior and the dependence structure across FAS ports \cite{ghadi2023copula,ghadi2023gaussian,tlebaldiyeva2023copula,r2024gaussian}. These works reveal that FAS channels can exhibit highly non-Gaussian, spatially correlated statistics that are well captured by appropriate copulas, and that the effective rank of the correlation matrix, shaped by the physical aperture and scattering geometry, is a key determinant of performance. Yet, this copula machinery has mainly been used in off-line analysis, not as a component in a real-time inference engine for port selection.

Beyond FAS, we have witnessed a rapid uptake of generative models for channel estimation and signal detection, including diffusion models and deep JSCC \cite{zappone2019dl_wireless,shen2021gnn_rrm,wu2024cddm,kurka2020deepjsccf}. A particularly relevant class of models learns high-dimensional joint distributions via attention and copula-inspired normalizing flows \cite{drouin2022tactis}. Such models naturally fit the FAS setting, in which we observe only a subset of spatio-temporal samples and wish to infer the joint behavior of the unobserved ones.

Despite these converging trends, there is, to the best of our knowledge, no existing work that directly tackles \emph{symbol-level interference imputation and port selection in the fast FAMA problem} using a generative, copula-based time-series model. This is the gap that the present paper aims to fill.

\vspace{-2mm}
\subsection{Contributions}
In this paper, we propose a \emph{copula-aided FAMA} framework that treats port selection as a probabilistic inference problem. Given only partial observations of the desired channel $\{h_k\}$ and received signal $\{r_k\}$ on a small subset of ports, we infer the posterior distribution of the interference $\{I_k\}$ and SINR over all FAS ports, and then apply selection and combining policies based on these inferred quantities.

Our main contributions are summarized as follows.
\begin{itemize}
\item We formalize the symbol-level fast FAMA port selection problem as a joint imputation problem over the complex-valued triplet $(r_k,h_k,I_k)$ at each port $k$. For each symbol, only the desired channel and received signal are observed on a subset of $M(<K)$ ports, and the goal is to infer the instantaneous interference and SINR on all $K$ ports. We consider both one-dimensional (1D) and two-dimensional (2D) FAS geometries, and both rich- and finite-scattering channel models with low effective rank.
\item We construct an attention copula time-series model tailored to FAS interference fields. The model operates on a two-channel representation (real and imaginary parts) of a length-$3K$ time series per symbol, corresponding to $(r_k,h_k,I_k)$ across ports. In particular, we design training masks that mimic practical reception constraints, so that the model learns to interpolate unobserved ports given only partial $(r,h)$ observations.
\item Crucially, we evaluate the proposed scheme under both rich and finite scattering, and for both 1D and 2D FAS layouts. Our results show that with only a modest number of reception ports $M \ll K$,\footnote{When only one RF chain is available, the FAS operates $M$ times faster than the symbol rate to take the $M$ signal samples per symbol duration.} the proposed copula-aided FAMA closely tracks an oracle that knows the true per-port SINR. The ability to operate with a tiny receiving set suggests that fast or near-fast FAMA may be feasible even with realistic switching and pilot constraints.
\end{itemize}

The remainder of this paper is organized as follows. Section~\ref{sec:system-model} introduces the system models for 1D and 2D FAS configurations under rich and finite scattering, together with the partial observation structure. Section~\ref{sec:full-hr-attentional-copula} then presents the proposed attentional-copula learning framework under full CSI and received signal observability, while Section~\ref{sec:general-attentional-copula} extends it to the general partially observed setting and also formulates the resulting copula-aided FAMA scheme. Section~\ref{sec:simulation_results} provides numerical results on interference imputation accuracy and rate performance. Finally, we discuss some practical implications in Section \ref{sec:practice} and conclude the paper in Section~\ref{sec:conclusion}.

\vspace{-2mm}
\section{System Model}\label{sec:system-model}
Consider a situation where a BS with multiple fixed-position antennas (FPAs) communicates to $U$ UTs on the same channel-use, without using precoding. Thus, without loss of generality, it is assumed that each FPA of the BS transmits to one UT. This model can also be interpreted as an interference channel for D2D communication as no prior centralized optimization is performed. Each UT is equipped with a FAS providing $K$ candidate antenna positions (referred to as `ports') inside a given aperture. By reconfiguring its active port on a fast time scale, each UT can exploit spatial fluctuations of interference and noise based on the fast FAMA principle. Unless otherwise stated, we shall focus on a representative UT, indexed by $u$, and drop the index when no ambiguity arises.

\vspace{-2mm}
\subsection{FAS Geometry}
Each UT has $K$ discrete candidate ports located in either a 1D or 2D aperture. Let $D\in\{1,2\}$ denote the FAS dimension and we have the set of port locations
\begin{equation}
\mathcal{P} \triangleq \{\mathbf{p}_k \in \mathbb{R}^D : k=1,\ldots,K\}.
\end{equation}

In the 1D case with a linear FAS with normalized aperture length $W$ (in wavelengths), the $K$ ports are uniformly placed along a segment of length $W\lambda$, where $\lambda$ is the carrier wavelength. The $k$-th port position is given by
\begin{equation}
\mathbf{p}_k = (x_k),~\mbox{where }x_k = \frac{k-1}{K-1} W\lambda,~k=1,\ldots,K.
\end{equation}

For a 2D FAS, we consider a rectangular aperture of size $W_x\lambda \times W_y\lambda$ populated by $K_x \times K_y$ candidate ports, with $K=K_xK_y$. Let $k \equiv (i,j)$ index the port at row $i$ and column $j$, where $i\in\{1,\ldots,K_x\}$ and $j\in\{1,\ldots,K_y\}$. Its Cartesian coordinates are expressed as
\begin{equation}
\mathbf{p}_{i,j}=\begin{bmatrix}
x_i\\
y_j
\end{bmatrix},~\mbox{where }
\left\{\begin{aligned}
x_i&=\frac{i-1}{K_x-1} W_x\lambda,\\
y_j&=\frac{j-1}{K_y-1} W_y\lambda.
\end{aligned}\right.
\end{equation}
%When convenient, we flatten $(i,j)$ into a single index $k\in\{1,\ldots,K\}$ and keep the mapping $k \leftrightarrow \mathbf{p}_k$ implicit.

\vspace{-2mm}
\subsection{Signal Model}
Let $s_u[n]\in\mathbb{C}$ represent the symbol transmitted towards UT~$u$ at symbol time $n$, with $\mathbb{E}[|s_u[n]|^2]=P_s$ and $\{s_u[n]\}$ independent across UTs and time. For notational simplicity we consider a single symbol interval and drop the time index.

Recalling from (\ref{eq:rk0}), we have the received signal at the $k$-th port of the UT concerned
\begin{equation}\label{eq:per-port-signal}
r_k = h_k s_u + I_k + \eta_k,
\end{equation}
where the complex channel from the BS to the $k$-th port of the UT is denoted by $h_k\in\mathbb{C}$ (desired channel), and $I_k=\sum_{\bar{u}\ne u}h_k^{(\bar{u})}s_{\bar{u}}$ is the data-dependent aggregate interference at port $k$, and $\eta_k \sim \mathcal{CN}(0,\sigma_\eta^2)$ is spatially and temporally white additive noise. Collecting all ports into vectors, we have
\begin{equation}
\left\{\begin{aligned}
\mathbf{r} &\triangleq [r_1,\ldots,r_K]^{\mathsf{T}},\\
\mathbf{h} &\triangleq [h_1,\ldots,h_K]^{\mathsf{T}},\\
\mathbf{I} &\triangleq [I_1,\ldots,I_K]^{\mathsf{T}},\\
\boldsymbol{\eta} &\triangleq [\eta_1,\ldots,\eta_K]^{\mathsf{T}}.
\end{aligned}\right.
\end{equation}
Hence, we have the per-symbol FAS snapshot
\begin{equation}\label{eq:vector-signal-model}
\mathbf{r} = \mathbf{h}s_u + \mathbf{I} + \boldsymbol{\eta}.
\end{equation}
The instantaneous SINR at port $k$, a.k.a.~the signal-interference split after ignoring the noise with interference the dominant factor (in a per-symbol sense), is defined as
\begin{equation}\label{eq:sinr-definition}
\bar{\gamma}_k\triangleq\frac{|h_k|^2 |s_u|^2}{|I_k|^2}.
\end{equation}
Note that no averaging in (\ref{eq:sinr-definition}) is taken place, hence regarded as {\em instantaneous}. Fast FAMA exploits the spatial variability of $\{\bar{\gamma}_k\}$ across ports by activating the `sweet spot' in which $|h_k|$ is strong and $|I_k|$ is weak. As $|s_u|$ is same regardless of which port is selected, the port selection process is based on the normalized instantaneous SINR
\begin{equation}\label{eq:nsinr}
\gamma_k\triangleq\frac{\bar{\gamma}_k}{|s_u|^2}=\frac{|h_k|^2}{|I_k|^2}.
\end{equation}

\vspace{-2mm}
\subsection{Channel Models}\label{subsec:channel-models}
We here describe the small-scale fading models adopted for the desired channel $\mathbf{h}$ and the interfering channels $\{h_k^{(\bar u)}\}$. Following the literature, we consider two canonical cases.

\subsubsection{Rich-scattering Gaussian model}
In rich scattering, the channel across FAS ports is well approximated by a zero-mean corrected Gaussian vector whose covariance depends only on spatial separations. For the desired link, we write
\begin{equation}
\mathbf{h} \sim \mathcal{CN}(\mathbf{0}, \Omega_h \mathbf{R}^{(D)}),
\end{equation}
where $\Omega_h$ captures large-scale effects and $\mathbf{R}^{(D)}$ is the normalized spatial correlation matrix.

For a 1D FAS with port locations $\{x_k\}$ as above, the $(k,\ell)$-th entry of $\mathbf{R}^{(1\mathrm{D})}\in\mathbb{C}^{K\times K}$ is
\begin{equation}\label{eq:J0-correlation-1d-ojcoms}
[\mathbf{R}^{(1\mathrm{D})}]_{k,\ell}
=J_0\!\left(2\pi \frac{|x_k - x_\ell|}{\lambda}\right)
= J_0\!\left(2\pi W \frac{|k-\ell|}{K-1}\right),
\end{equation}
with $J_0(\cdot)$ the zeroth-order Bessel function of the first kind.

For a 2D FAS, we have
\begin{equation}
\mathbf{h} \sim \mathcal{CN}(\mathbf{0}, \Omega_h \mathbf{R}^{(2\mathrm{D})}),
\end{equation}
where the $(m,n)$-th element of $\mathbf{R}^{(2\mathrm{D})}$ depends only on the distance between port locations $\mathbf{p}_m$ and $\mathbf{p}_n$, and 
\begin{equation}\label{eq:J0-correlation-2d-ojcoms}
[\mathbf{R}^{(2\mathrm{D})}]_{m,n}
= j_0\!\left(2\pi \frac{\|\mathbf{p}_m - \mathbf{p}_n\|}{\lambda}\right),
\end{equation}
where $j_0(\cdot)$ is the (spherical) Bessel function of order zero. Equivalently, if $m\equiv(m_1,m_2)$ and $n\equiv(n_1,n_2)$ index the ports on a $K_x\times K_y$ grid with normalized aperture sizes $W_x,W_y$, \eqref{eq:J0-correlation-2d-ojcoms} reduces to the standard Jakes' correlation. The interfering channels $\{h_k^{(\bar u)}\}$ are modeled in the same way.

\subsubsection{Finite-scattering Rician model}
To capture the characteristics of environments dominated by a finite number of scattering clusters (e.g., millimeter-wave (mmWave) propagation), we adopt a Rician finite-scattering model where the effective channel vector $\mathbf{h}\in\mathbb{C}^{K}$ is written as
\begin{multline}\label{eq:finite-scattering-general}
\mathbf{h}= \sqrt{\frac{K_R}{K_R+1}}\,e^{j\delta}\,
\mathbf{a}_0(\theta_0,\varphi_0) \\
+ \sqrt{\frac{1}{K_R+1}}\frac{1}{\sqrt{N_p}}\sum_{\ell=1}^{N_p}
\kappa_\ell\,\mathbf{a}_\ell(\theta_\ell,\varphi_\ell),
\end{multline}
where $K_R$ is the Rician factor, $\delta$ is the deterministic line-of-sight (LoS) phase, $N_p$ is the number of non-LoS paths, $\kappa_\ell\sim\mathcal{CN}(0,\beta_\ell)$ is the complex gain of path $\ell$, and $\mathbf{a}_\ell(\theta_\ell,\varphi_\ell)\in\mathbb{C}^K$ is the steering vector across the FAS ports for the azimuth and elevation angle-of-arrivals (AoAs) $(\theta_\ell,\varphi_\ell)$.

For a 2D FAS with $K_x$ and $K_y$ ports along the horizontal and vertical axes and spacings
\begin{equation}
\Delta_d = \frac{W_d\lambda}{K_d - 1},~d\in\{x,y\},
\end{equation}
the port $(n_1,n_2)$ is located at
\begin{equation}
\mathbf{p}_{n_1,n_2}
= \begin{bmatrix}
(n_1-1)\Delta_x \\
(n_2-1)\Delta_y
\end{bmatrix}.
\end{equation}
For path $\ell$ with AoAs $(\theta_\ell,\varphi_\ell)$, the excess propagation distance between port $(n_1,n_2)$ and the reference port $(1,1)$ is
\begin{equation}\label{eq:prop-diff-2d}
S_\ell(n_1,n_2)= (n_1-1)\Delta_x \sin\theta_\ell \cos\varphi_\ell+ (n_2-1)\Delta_y \cos\theta_\ell,
\end{equation}
and the corresponding steering vector is
\begin{equation}\label{eq:steering-2d}
\mathbf{a}_\ell(\theta_\ell,\varphi_\ell)=\left[1,\,
e^{-j\frac{2\pi}{\lambda}S_\ell(1,2)},\,
\ldots,\,
e^{-j\frac{2\pi}{\lambda}S_\ell(K_x,K_y)}
\right]^{\mathsf{T}}.
\end{equation}

The 1D FAS is obtained as the special case $K_y=1$, $W_y=0$. Then $k\equiv n_1\in\{1,\ldots,K_x\}$, with $x_k=(k-1)\Delta_x$ and
\begin{equation}\label{eq:prop-diff-1d}
S_\ell(k)= (k-1)\Delta_x \sin\theta_\ell \cos\varphi_\ell,
\end{equation}
so that the 1D steering vector becomes
\begin{equation}\label{eq:steering-1d}
\mathbf{a}_\ell(\theta_\ell,\varphi_\ell)= \left[
1,\,
e^{-j\frac{2\pi}{\lambda}S_\ell(2)},\,
\ldots,\,
e^{-j\frac{2\pi}{\lambda}S_\ell(K_x)}
\right]^{\mathsf{T}}.
\end{equation}
Under \eqref{eq:finite-scattering-general} with independent $\{\kappa_\ell\}$, the channel covariance matrix can be found as
\begin{equation}\label{eq:R-finite-2d-ojcoms}
\mathbf{R}^{(2\mathrm{D})}_{\text{finite}}
\triangleq \mathbb{E}[\mathbf{h}\mathbf{h}^{\mathsf{H}}]
= \frac{K_R}{K_R+1}\,\mathbf{a}_0\mathbf{a}_0^{\mathsf{H}}+ \frac{1}{K_R+1}\,\frac{1}{N_p}\sum_{\ell=1}^{N_p}\beta_\ell\,
\mathbf{a}_\ell\mathbf{a}_\ell^{\mathsf{H}},
\end{equation}
whose rank is at most $1+N_p$, reflecting the limited number of scattering clusters. The interfering links can be modeled in the same way. Note that we always use the same model for both the desired and interfering links.

\vspace{-3mm}
\subsection{Fast FAMA and Ideal Port Selection}\label{subsec:fast-fama}
In the classical fast FAMA paradigm, the UT is assumed to have access to the instantaneous received signal $\{r_k\}_{k=1}^K$ and the instantaneous interference signal $\{I_k\}_{k=1}^K$ at every port during each symbol interval. As such, the UT can conceptually evaluate the per-port SINRs $\{\gamma_k\}$ in
\eqref{eq:nsinr} and solve
\begin{equation}\label{eq:ideal-port-selection}
k^\star= \arg\max_{1\le k\le K} \gamma_k.
\end{equation}
%or a set of ports for further combining. This symbol-level port selection exploits deep fades of the aggregate interference $I_k$ and constructive fluctuations of the desired channel $h_k$, enabling strong interference mitigation without transmitter-side CSI or centralized coordination.

As explained before, fully observing $\{r_k\}_{k=1}^K$ at symbol rate is challenging due to RF-chain limitations. Additionally, $\gamma_k$ depends on the data-dependent interference knowledge $|I_k|^2$, which is not directly observable. Consequently, a practical fast FAMA scheme must operate with only partial observations of the FAS snapshot and incomplete knowledge of $\{I_k\}$.

\vspace{-3mm}
\subsection{Interference Field and Learning Problem}\label{subsec:learning-problem}
To formalize our problem, we introduce the per-port triplet 
\begin{equation}
\mathbf{z}_k \triangleq \left(r_k,h_k, I_k\right) \in \mathbb{C}^3,
\end{equation}
and the stacked random vector
\begin{equation}
    \mathbf{z}
    \triangleq
    [\mathbf{z}_1^{\mathsf{T}},\ldots,\mathbf{z}_K^{\mathsf{T}}]^{\mathsf{T}}
    \in \mathbb{C}^{3K},
\end{equation}
which captures the joint statistics of the desired channel, the aggregate interference, and the received signal across all ports.

We assume that the UT can directly observe only a subset $\mathcal{M}\subset\{1,\ldots,K\}$ of ports per symbol, with $|\mathcal{M}| = M \ll K$. For $k\in\mathcal{M}$, we further assume that channel estimation provides $h_k$ (from pilots) and that the received signal sample $r_k$ is observed. For the remaining ports $k\notin\mathcal{M}$, the quantities $(h_k,r_k,I_k)$ are unobserved. We denote
\begin{align}
\mathbf{z}_{\text{obs}}&\triangleq \{\mathbf{z}_k: k\in\mathcal{M}\},\\
\mathbf{z}_{\text{miss}}&\triangleq \{\mathbf{z}_k: k\notin\mathcal{M}\}.
\end{align}

The key objective of this paper is to learn the conditional law of the unobserved interference field $\{I_k\}_{k=1}^K$ (and more generally, $\mathbf{z}_{\text{miss}}$) given the partial observations $\mathbf{z}_{\text{obs}}$. Formally, we seek a generative model that approximates
\begin{equation}
p\left(\mathbf{z}_{\text{miss}}\mid\mathbf{z}_{\text{obs}}\right).
\end{equation}
Such a model allows the UT to impute the interference power profile $\{|I_k|^2\}$ across all ports from only a small subset of observed ports, enabling fast FAMA decisions.

To this end, we develop an attentional copula based generative model tailored to the structured dependence in $\mathbf{z}$, and demonstrate that it can accurately reconstruct the interference field under both rich and finite scattering cases.

\vspace{-2mm}
\section{Attentional-Copula Learning with\\Full $(\mathbf{r},\mathbf{h})$ Observability}\label{sec:full-hr-attentional-copula}
Here, we present a simplified version of the learning problem in Section~\ref{subsec:learning-problem}, where the UT is assumed to have symbol-wise access to the desired channel vector $\mathbf{h}$ and the received signal vector $\mathbf{r}$ across all $K$ ports, while the interference field $\mathbf{I}$ remains unobserved. Our goal is to learn the conditional law
\begin{equation}
p(\mathbf{I} \mid \mathbf{r},\mathbf{h}),
\end{equation}
and draw approximate posterior samples of $\mathbf{I}$ given an observed snapshot $(\mathbf{r},\mathbf{h})$. Crucially, the desired symbol $s_u$ and the interfering symbols $\{s_{\bar u}\}$ are never estimated explicitly.%, their effect is integrated out through the data distribution.

The learning problem is tackled with a two-stage attentional copula mode, adapted to the FAS snapshot structure and the FAMA signal model. In the following, we start with basic preliminaries on copulas and transformer-based conditional modelling, then describe the port-major representation and the two-stage architecture for the interference-only task.

\vspace{-2mm}
\subsection{Copula and Transformer Preliminaries}\label{subsec:copula-transformer-prelim}
Let $\mathbf{X} = (X_1,\ldots,X_D)$ denote a continuous random vector with marginal cumulative distribution functions (CDFs) $F_i(x) \triangleq p(X_i \le x)$, $i=1,\ldots,D$. Sklar's theorem asserts the existence of a copula $C:[0,1]^D \to [0,1]$ such that
\begin{equation}
p(X_1 \le x_1,\ldots,X_D \le x_D)= C\left(F_1(x_1),\ldots,F_D(x_D)\right).
\end{equation}
If the joint density $f_{\mathbf{X}}$ exists and the marginals admit densities $f_i$, then the copula admits a density $c$ on $[0,1]^D$ and
\begin{equation}
f_{\mathbf{X}}(\mathbf{x})=c\left(F_1(x_1),\ldots,F_D(x_D)\right)\prod_{i=1}^D f_i(x_i).
\end{equation}

Equivalently, by letting $U_i = F_i(X_i)$, we have $\mathbf{U}=(U_1,\ldots,U_D) \in [0,1]^D$ with joint density $c$, and each $U_i$ is standard uniform. The copula thus captures all dependence between coordinates, decoupled from the marginals.

In parametric form, let $\phi$ denote the parameters of the marginal CDFs $\{F_{i,\phi}\}_{i=1}^D$ and $\theta$ the parameters of a copula density $c_\theta$. The resulting joint model can be written as
\begin{equation}
g_{\theta,\phi}(\mathbf{x})\triangleq c_\theta\left(F_{1,\phi}(x_1),\ldots,F_{D,\phi}(x_D)\right)\prod_{i=1}^D f_{i,\phi}(x_i),
\end{equation}
where $f_{i,\phi} = \partial_x F_{i,\phi}$. The log-density decomposes as 
\begin{equation}
\log g_{\theta,\phi}(\mathbf{x})= \log c_\theta(\mathbf{u})+\sum_{i=1}^D \log f_{i,\phi}(x_i),
\end{equation}
where we have ${\bf u}=(u_1,\dots,u_D)$ in which $u_i = F_{i,\phi}(x_i)$. This clearly illustrates the separation between marginal modelling and dependence modelling.

In our setting, $\mathbf{X}$ will represent a real-valued encoding of the FAS snapshot $(\mathbf{r},\mathbf{h},\mathbf{I})$ across ports in terms of their real and imaginary parts. We model the marginals via tractable monotone normalizing flows and the copula via a transformer-parametrized density, referred to as an \emph{attentional copula}.

For conditional modelling, consider a partition of indices $\{1,\ldots,D\} = \mathcal{H} \cup \mathcal{P}$ into observed coordinates (historicals) and prediction targets. We write $\mathbf{X}_{\mathcal{H}}$ and $\mathbf{X}_{\mathcal{P}}$ for the corresponding subvectors, and likewise $\mathbf{U}_{\mathcal{H}}$ and $\mathbf{U}_{\mathcal{P}}$. The conditional density $p(\mathbf{x}_{\mathcal{P}} \mid \mathbf{x}_{\mathcal{H}})$ can be factorized as
\begin{equation}
p(\mathbf{x}_{\mathcal{P}} \mid \mathbf{x}_{\mathcal{H}})=c_{\mathcal{P} | \mathcal{H}}\left(\mathbf{u}_{\mathcal{P}} \mid \mathbf{u}_{\mathcal{H}},\mathbf{x}_{\mathcal{H}}\right)\prod_{i\in\mathcal{P}} f_i(x_i),
\end{equation}
in which $c_{\mathcal{P}\mid\mathcal{H}}$ is a \emph{conditional copula density}. In our implementation, $c_{\mathcal{P}\mid\mathcal{H}}$ is parameterized by a transformer encoder-decoder operating on a set of tokens associated with series/time indices, with self-attention over the historical tokens and cross-attention from prediction tokens to historical tokens. Consequently, the transformer learns to map observed tokens $(\mathbf{u}_{\mathcal{H}}, \mathbf{x}_{\mathcal{H}})$ into rich context embeddings that define a flexible conditional copula for the prediction tokens.

\vspace{-2mm}
\subsection{Port-Major Representation and Snapshot Process}\label{subsec:port-major-process}
At a given symbol time, for each sample, we generate
%let $s_u$ denote the desired QPSK symbol and $s_{\bar u}$ the QPSK symbols of $U$ independent co-channel interferers, all of unit average power. 
\begin{align}
\mathbf{h}&= \mathbf{L}\mathbf{z}_0, &\mathbf{z}_0 &\sim \mathcal{CN}(\mathbf{0},\mathbf{I}_K),\\
\mathbf{h}^{(\ell)}&= \mathbf{L}\mathbf{z}_\ell, &\mathbf{z}_\ell &\sim \mathcal{CN}(\mathbf{0},\mathbf{I}_K),~\ell=1,\ldots,U,
\end{align}
where $\mathbf{L}$ is a Cholesky factor of the spatial correlation matrix $\mathbf{R}$. The aggregate interference across ports in vector form is
\begin{equation}
\mathbf{I} = \sum_{\ell=1\atop \ell\ne u}^{U} \mathbf{h}^{(\ell)} s_\ell.
\end{equation}
With the additive noise vector $\boldsymbol{\eta}\sim\mathcal{CN}(\mathbf{0}, \sigma_\eta^2\mathbf{I}_K)$, the received vector satisfies
\begin{equation}
\mathbf{r} = \mathbf{h} s_u + \mathbf{I} + \boldsymbol{\eta}.
\end{equation}

We encode a snapshot $(\mathbf{r},\mathbf{h},\mathbf{I})$ as a real-valued process $\mathbf{X} \in \mathbb{R}^{2\times T}$ with $T=3K$ `time' steps and two series for real and imaginary parts. For port index $k\in\{1,\ldots,K\}$ we define
\begin{equation}
\left\{\begin{aligned}
t_r(k) &\triangleq 3k-2,\\
t_h(k) &\triangleq 3k-1,\\
t_I(k) &\triangleq 3k.
\end{aligned}\right.
\end{equation}
The entries of $\mathbf{X}$ are then arranged as
\begin{equation}
\left\{\begin{aligned}
X_{1,t_r(k)} &= \Re\{r_k\},\\
X_{1,t_h(k)} &= \Re\{h_k\},\\
X_{1,t_I(k)} &= \Re\{I_k\},\\
X_{2,t_r(k)} &= \Im\{r_k\},\\
X_{2,t_h(k)} &= \Im\{h_k\},\\
X_{2,t_I(k)} &= \Im\{I_k\}.
\end{aligned}\right.
\end{equation}
Therefore, each training example is a length-$T$ bivariate time series, with structured groups of three consecutive time indices corresponding to one physical port.

For some derivations, we find it convenient to flatten $\mathbf{X}$ into a vector $\tilde{\mathbf{X}}\in\mathbb{R}^D$ with
\begin{equation}
D \triangleq 2T = 6K,
\end{equation}
using a fixed bijection $i \leftrightarrow (d,t)$ between scalar indices $i\in\{1,\ldots,D\}$ and pairs $(d,t)$ with $d\in\{1,2\}$, $t\in\{1,\ldots,T\}$. The interference-only task will be defined on index sets
\begin{align}
\mathcal{T}_r &\triangleq \{t_r(k): 1\le k\le K\},\\
\mathcal{T}_h &\triangleq \{t_h(k): 1\le k\le K\},\\
\mathcal{T}_I &\triangleq \{t_I(k): 1\le k\le K\}.
\end{align}

\vspace{-2mm}
\subsection{Conditional Interference Field $p(\mathbf{I}\mid\mathbf{r},\mathbf{h})$}\label{subsec:conditional-I-only}
In the simplified setting considered here, we assume that on each symbol the UT has accurate instantaneous estimates of the desired channel and received signal at every port, i.e.,
\begin{equation}
\mathbf{r},\mathbf{h}~\text{(fully observed)},~\mbox{but }\mathbf{I}~\text{(unobserved)}.
\end{equation}
In the real-valued process $\mathbf{X}$, we declare the index sets
\begin{align}
\mathcal{H} &\triangleq\{(d,t): d\in\{1,2\}, t\in \mathcal{T}_r\cup\mathcal{T}_h\},\\
\mathcal{P} &\triangleq \{(d,t): d\in\{1,2\}, t\in \mathcal{T}_I\},
\end{align}
so that $\mathcal{H}$ contains all real and imaginary parts of $(\mathbf{r}, \mathbf{h})$ and $\mathcal{P}$ contains all real and imaginary parts of $\mathbf{I}$. The model is trained to approximate the conditional density
\begin{equation}
p_{\theta,\phi}(\mathbf{X}_\mathcal{P} \mid \mathbf{X}_\mathcal{H})\approx p(\mathbf{X}_\mathcal{P} \mid \mathbf{X}_\mathcal{H}),
\end{equation}
in which $\phi$ denotes the marginal-flow parameters and $\theta$ denotes the copula parameters.

More concretely, let $X_i$ be the scalar at index $i=(d,t)\in\mathcal{H}\cup\mathcal{P}$ and $F_{i,\phi}$ its parametric marginal CDF.\footnote{For convenience, we will keep using 1D indexing to simplify our notations and convert it back to 2D when needed, e.g., from (\ref{eq:1Dx}) to (\ref{eq:2Di}).} We define
\begin{equation}
U_i = F_{i,\phi}(X_i),~i\in\mathcal{H}\cup\mathcal{P},
\end{equation}
and model the conditional copula density
\begin{equation}\label{eq:a-copula}
c_\theta(\mathbf{u}_\mathcal{P} \mid \mathbf{u}_\mathcal{H},\mathbf{x}_\mathcal{H}).
\end{equation}
By doing so, the resulting conditional density on the original scale can be constructed as
\begin{equation}
p_{\theta,\phi}(\mathbf{x}_\mathcal{P} \mid \mathbf{x}_\mathcal{H})= c_\theta\left(\mathbf{u}_\mathcal{P} \mid\mathbf{u}_\mathcal{H},\mathbf{x}_\mathcal{H}\right)
\prod_{i\in\mathcal{P}} f_{i,\phi}(x_i),
\end{equation}
where $f_{i,\phi} = F_{i,\phi}'$ are the marginal densities.

At test time, given an observed snapshot $(\mathbf{r},\mathbf{h})$, 
\begin{itemize}
\item Construct $\mathbf{x}_\mathcal{H}$; 
\item Compute the corresponding $\mathbf{u}_\mathcal{H}$, and 
\item Sample $\mathbf{u}_\mathcal{P}$ from the attentional copula (\ref{eq:a-copula}). 
\end{itemize}
The interference field samples are then obtained via
\begin{equation}\label{eq:1Dx}
\hat{X}_i^{(m)} = F_{i,\phi}^{-1}\left(U_i^{(m)}\right),~i\in\mathcal{P}~\mbox{and } m=1,\ldots,M,
\end{equation}
and aggregated into complex interference samples 
\begin{equation}\label{eq:2Di}
\hat{I}_k^{(m)}=\hat{X}_{1,t_I(k)}^{(m)} + j \hat{X}_{2,t_I(k)}^{(m)}.
\end{equation}

\vspace{-3mm}
\subsection{Two-Stage Attentional Copula Architecture}\label{subsec:two-stage-architecture}
The architecture follows a two-stage design.

\subsubsection{Stage~1: Marginal normalizing flows}
Stage~1 fits parametric univariate marginals for each scalar coordinate $X_i$. We employ a deep sigmoidal flow (DSF) family for $F_{i,\phi}$, which can be written in terms of an invertible transformation $T_{i,\phi}:\mathbb{R}\to\mathbb{R}$ and the standard normal CDF $\Phi$:
\begin{equation}
U_i = F_{i,\phi}(X_i)= \Phi\left(T_{i,\phi}(X_i)\right),~i\in\mathcal{H}\cup\mathcal{P}.
\end{equation}
Each $T_{i,\phi}$ is realized by a small monotone neural network (a composition of affine layers and strictly monotone pointwise transformations) whose parameters are shared across series but modulated by learned embeddings of the indices $d$ and $t$.

The marginal $f_{i,\phi}$ follows from the change-of-variables
\begin{equation}
f_{i,\phi}(x)= \varphi\left(T_{i,\phi}(x)\right)\left|\frac{\mathrm{d}}{\mathrm{d}x} T_{i,\phi}(x)\right|,
\end{equation}
where $\varphi(\cdot)$ denotes the standard normal density.

Given $N$ independent and identically distributed (i.i.d.) snapshots $\{\mathbf{X}^{(n)}\}_{n=1}^N$ drawn from the simulated distribution $p_{\text{sim}}(\mathbf{X})$, Stage~1 maximizes the factorized log-likelihood
\begin{equation}
\mathcal{L}_{\text{marg}}(\phi)= \frac{1}{N} \sum_{n=1}^N \sum_{i}\log f_{i,\phi}\left(X^{(n)}_i\right),
\end{equation}
which in the population limit corresponds to
\begin{equation}
\phi^\star\in \arg\max_{\phi}\mathbb{E}_{\mathbf{X}\sim p_{\text{sim}}}\left[\sum_{i=1}^D \log f_{i,\phi}(X_i)\right].
\end{equation}
In practice, the loss is computed as the negative mean log-Jacobian determinant of the marginal flows.

\subsubsection{Stage~2: Transformer-parameterized copula}
In Stage~2, the marginal parameters $\phi$ are frozen and a transformer-based copula model with parameters $\theta$ is trained. For each index
$i=(d,t)$, we build an input token
\begin{equation}
\mathbf{e}_i= \mathbf{e}_{\text{series}}(d)+ \mathbf{e}_{\text{time}}(t)+ \mathbf{e}_{\text{type}}(t),
\end{equation}
in which $\mathbf{e}_{\text{series}}$, $\mathbf{e}_{\text{time}}$, and $\mathbf{e}_{\text{type}}$ are learnable embeddings that encode the series index, the position along the FAS (port-major index), and the variable type ($r$, $h$, or $I$), respectively. The observed values are embedded through simple linear projections of $U_i$ or $X_i$ and added to $\mathbf{e}_i$.

The historical tokens $i\in\mathcal{H}$ are passed through a temporal transformer encoder (stack of self-attention and feedforward blocks) to produce context vectors $\mathbf{c}_i$. The prediction tokens $i\in\mathcal{P}$ are initialized with their positional/type embeddings and pointwise noise, then processed by a transformer decoder with cross-attention over $\{\mathbf{c}_i\}_{i\in\mathcal{H}}$. The decoder outputs parameter vectors that specify, for each prediction coordinate $i\in\mathcal{P}$, a univariate conditional density on $U_i \in (0,1)$. This conditional density is realized via a monotone flow on $(0,1)$, yielding the attentional copula $c_\theta$.

Formally, the copula factorization is written as
\begin{equation}
c_\theta(\mathbf{u}_\mathcal{P} \mid \mathbf{u}_\mathcal{H},\mathbf{x}_\mathcal{H})= \prod_{i\in\mathcal{P}} c_{\theta,i}\left(u_i \mid \mathbf{u}_\mathcal{H},\mathbf{x}_\mathcal{H}\right),
\end{equation}
where each $c_{\theta,i}$ is the density induced by the decoder output for coordinate $i$.

\vspace{-2mm}
\subsection{Training Objective and Evaluation Metric}\label{subsec:training-eval-I-only}
The full training procedure alternates the two stages
\begin{itemize}
\item \textbf{Stage~1:} Optimize $\phi$ to maximize $\mathcal{L}_{\text{marg}}(\phi)$ over i.i.d.\ snapshots $\mathbf{X}$.
\item \textbf{Stage~2:} Freeze $\phi$, and for each batch construct $(\mathcal{H},\mathcal{P})$ as in Section~\ref{subsec:conditional-I-only} (observing all $(\mathbf{r},\mathbf{h})$ coordinates and predicting all interference coordinates ${\bf I}$). Optimize $\theta$ to maximize the conditional log-likelihood
\begin{equation}
\mathcal{L}_{\text{cop}}(\theta)= \frac{1}{N}\sum_{n=1}^N\sum_{i\in\mathcal{P}}\log c_{\theta,i}\left(u_i^{(n)} \mid\mathbf{u}_\mathcal{H}^{(n)},\mathbf{x}_\mathcal{H}^{(n)}\right),
\end{equation}
or equivalently minimizing $-\mathcal{L}_{\text{cop}}(\theta)$.
\end{itemize}

In the population limit, this Stage~2 objective can be viewed as minimizing an expected conditional Kullback-Leibler (KL) divergence. Let $q(\mathbf{u}_\mathcal{P} \mid \mathbf{u}_\mathcal{H},\mathbf{x}_\mathcal{H})$ denote the true conditional copula density. Then we have
\begin{align}
    \theta^\star
    &\in \arg\max_\theta
        \mathbb{E}_{\mathbf{X}\sim p_{\text{sim}}}
        \left[\log c_\theta(\mathbf{U}_\mathcal{P}
               \mid \mathbf{U}_\mathcal{H},\mathbf{X}_\mathcal{H})\right]
    \nonumber\\
    &=
    \arg\min_\theta
    \mathbb{E}_{\mathbf{X}\sim p_{\text{sim}}}
    \left[
      \mathrm{KL}\left(
        q(\cdot \mid \mathbf{U}_\mathcal{H},\mathbf{X}_\mathcal{H})
        \,\big\|\,
        c_\theta(\cdot \mid \mathbf{U}_\mathcal{H},\mathbf{X}_\mathcal{H})
      \right)
    \right]
    \nonumber\\
    &\quad + \text{constant},
\end{align}
where the additive constant does not depend on $\theta$.

\vspace{-2mm}
\section{General Attentional-Copula FAMA with Partial Observations}\label{sec:general-attentional-copula}
We now return to the realistic setting of Section~\ref{subsec:learning-problem}, where the UT can observe only a strict subset of ports per symbol and cannot access the interference field. In this situation, fast FAMA cannot rely on direct evaluation of the per-port SINRs $\{\gamma_k\}$, and must instead infer the full interference field and the unobserved channel coefficients from limited observations.

We now extend the attentional copula framework to arbitrary partially observed snapshots, learning the conditional law
\begin{equation}
p_{\theta,\phi}\left(\mathbf{z}_{\text{miss}} \mid \mathbf{z}_{\text{obs}}\right)\approx p\left(\mathbf{z}_{\text{miss}} \mid \mathbf{z}_{\text{obs}}\right),
\end{equation}
for any feasible observation pattern of the triplets $\mathbf{z}_k = (r_k,h_k,I_k)$ across ports. This allows the UT to impute the interference profile and approximate the fast-FAMA decision \eqref{eq:ideal-port-selection} using only a small number of signal measurements.

\vspace{-2mm}
\subsection{Observation Model and Masked Conditional Law}\label{subsec:observation-mask}
Recall the real-valued encoding of a snapshot as $\mathbf{X}\in\mathbb{R}^{2\times T}$, $T=3K$, with indices $i\leftrightarrow(d,t)$ defined as in Section~\ref{subsec:port-major-process}. We introduce a random binary mask
\begin{equation}
\mathbf{M} = (M_i)_{i=1}^{2T} \in \{0,1\}^{2T},
\end{equation}
indicating which scalar coordinates are directly observed at the UT. Given $\mathbf{M}$ and $\mathbf{X}$, we define
\begin{align}
\mathcal{H}(\mathbf{M})&\triangleq \{ i : M_i = 1\},\\
\mathcal{P}(\mathbf{M})&\triangleq \{ i : M_i = 0\},
\end{align}
so that $\mathbf{X}_{\mathcal{H}(\mathbf{M})}$ are observed and $\mathbf{X}_{\mathcal{P}(\mathbf{M})}$ are missing.

The partial-observation model is defined via a distribution $q(\mathbf{M})$ over feasible masks, capturing practical constraints.  At each symbol time, the UT randomly selects a subset $\mathcal{M}\subset\{1,\ldots,K\}$ with $|\mathcal{M}|=M\ll K$, and measures both $h_k$ (from pilots) and $r_k$ (direct reception) only for $k\in\mathcal{M}$. Then $\mathcal{H}(\mathbf{M})$ contains the real and imaginary parts of $h_k$ and $r_k$ for $k\in\mathcal{M}$, while all other coordinates are in $\mathcal{P}(\mathbf{M})$.

Given $\mathbf{M}$, we seek to model the conditional density
\begin{equation}
p(\mathbf{x}_{\mathcal{P}(\mathbf{M})}\mid \mathbf{x}_{\mathcal{H}(\mathbf{M})},\mathbf{M}),
\end{equation}
both for training (with full simulated snapshots) and inference (where only $\mathbf{x}_{\mathcal{H}(\mathbf{M})}$ is available to the UT).

\vspace{-2mm}
\subsection{Masked Attentional Copula and Random-Mask Training}\label{subsec:masked-copula}
The marginals $F_{i,\phi}$ fitted in Stage~1 are unconditional and do not depend on the mask. Hence, the Stage-1 transformation to uniforms $U_i=F_{i,\phi}(X_i)$ still applies for any mask $\mathbf{M}$. The key change is in the copula, which must now handle arbitrary subsets of observed and missing coordinates.

We extend the attentional copula as follows. For each index $i=(d,t)$, we build token embeddings as in Section~\ref{subsec:two-stage-architecture}, and additionally include a binary mask embedding $\mathbf{e}_{\text{mask}}(M_i)$. The historical token set $\mathcal{H}(\mathbf{M})$ is passed through a transformer encoder to produce context vectors, while the prediction token set $\mathcal{P}(\mathbf{M})$ is fed to a cross-attention decoder. The decoder outputs, for each $i\in\mathcal{P}(\mathbf{M})$, the parameters of a conditional density on $U_i\in(0,1)$, resulting in the factorization
\begin{multline}
c_\theta\left(\mathbf{u}_{\mathcal{P}(\mathbf{M})}\mid \mathbf{u}_{\mathcal{H}(\mathbf{M})},\mathbf{x}_{\mathcal{H}(\mathbf{M})},\mathbf{M}\right)\\
= \prod_{i\in\mathcal{P}(\mathbf{M})}c_{\theta,i}\left(u_i \mid \mathbf{u}_{\mathcal{H}(\mathbf{M})},\mathbf{x}_{\mathcal{H}(\mathbf{M})},\mathbf{M}\right).
\end{multline}

Training proceeds by sampling both snapshots and masks:
\begin{align}
\mathbf{X} &\sim p_{\text{sim}}(\mathbf{X}), \\
\mathbf{M} &\sim q(\mathbf{M}),
\end{align}
where $p_{\text{sim}}$ is the simulation distribution induced by the FAS channels in Section~\ref{sec:system-model}. For each pair $(\mathbf{X},\mathbf{M})$, we compute the uniforms $U_i = F_{i,\phi}(X_i)$ and maximize %the conditional log-likelihood
\begin{multline}
\mathcal{L}_{\text{cop}}(\theta)=\\
\mathbb{E}_{{\mathbf{X}\sim p_{\text{sim}}\atop \mathbf{M}\sim q}}\Bigg[\sum_{i\in\mathcal{P}(\mathbf{M})}\log c_{\theta,i}\left(U_i \mid\mathbf{U}_{\mathcal{H}(\mathbf{M})},\mathbf{X}_{\mathcal{H}(\mathbf{M})},\mathbf{M}\right)\Bigg],
\end{multline}
which is estimated using Monte-Carlo simulations over mini-batches. As in the full observability case, this can be viewed as minimizing an expected conditional KL divergence
\begin{align}
\theta^\star&\in \arg\max_\theta \mathcal{L}_{\text{cop}}(\theta)\nonumber\\
&=\arg\min_\theta\mathbb{E}_{\mathbf{X}\sim p_{\text{sim}},\mathbf{M}\sim q}
    \Big[
      \mathrm{KL}\big(
        q(\cdot \mid
          \mathbf{U}_{\mathcal{H}(\mathbf{M})},
          \mathbf{X}_{\mathcal{H}(\mathbf{M})},\mathbf{M})
        \,\big\|\,
    \nonumber\\
    &\qquad
        c_\theta(\cdot \mid
          \mathbf{U}_{\mathcal{H}(\mathbf{M})},
          \mathbf{X}_{\mathcal{H}(\mathbf{M})},\mathbf{M})
      \big)
    \Big]+ \text{constant},
\end{align}
where $q$ denotes the true (unknown) conditional copula.

This masked training exposes the copula to a wide variety of observation patterns, enabling it to generalize to the specific hardware mask used at inference time. The training procedure is summarized in Algorithm \ref{alg:attentional_copula_fama}.

\vspace{-2mm}
\subsection{Interference Field Reconstruction and Fast FAMA}\label{subsec:reconstruction-fama}
At run time on each symbol, the UT operates to:
\begin{enumerate}
\item \textbf{Acquire partial observations.} Using its available RF chains and switching constraints, the UT probes a subset of ports $\mathcal{M}$ and obtains noisy estimates of $\{r_k,h_k\}_{k\in\mathcal{M}}$. These measurements define the observed coordinate set $\mathcal{H}(\mathbf{M})$ and corresponding values $\mathbf{x}_{\mathcal{H}(\mathbf{M})}$.
\item \textbf{Form posterior samples.} The UT transforms the observed coordinates to uniforms via $U_i = F_{i,\phi}(X_i)$ for $i\in\mathcal{H}(\mathbf{M})$, then repeatedly samples
\begin{equation}
\mathbf{u}_{\mathcal{P}(\mathbf{M})}^{(m)}\sim c_\theta\left(\cdot \mid\mathbf{u}_{\mathcal{H}(\mathbf{M})},\mathbf{x}_{\mathcal{H}(\mathbf{M})},\mathbf{M}\right),
~m=1,\ldots,M,
\end{equation}
and applies the inverse marginals $\hat{X}_i^{(m)} = F_{i,\phi}^{-1}(U_i^{(m)})$ for all $i\in\mathcal{P}(\mathbf{M})$.
\item \textbf{Reconstruct fields.} From the reconstructed scalar process $\hat{\mathbf{X}}^{(m)}$, we recover complex-valued samples $\hat{r}_k^{(m)}$, $\hat{h}_k^{(m)}$, and $\hat{I}_k^{(m)}$ for all ports $k=1,\ldots,K$, including those never probed in this symbol instance.
\end{enumerate}

\begin{remark}
Under rich-scattering Rayleigh channel conditions, the true joint field $(\mathbf{h},\mathbf{I})$ is (complex) Gaussian across ports, so the true copula is Gaussian as well. However, under finite-scattering models and structured interference, the dependence becomes markedly non-Gaussian, especially in 2D FAS geometries. The transformer-parametrized copula is sufficiently expressive to capture these high-order dependencies, and training on simulated snapshots generated according to the 1D and 2D rich and finite-scattering channel models allows the learned conditional law to faithfully reflect the physical channel.
\end{remark}

%\begin{figure}
\begin{minipage}{.9\linewidth}
\begin{algorithm}[H]
\caption{Two-Stage Attentional-Copula FAMA Training and Inference}\label{alg:attentional_copula_fama}
%\resizebox{.95\linewidth}{!}{
\begin{footnotesize}
\begin{algorithmic}[1]
\Require Synthetic snapshot simulator \( p_{\text{sim}}(\mathbf{X}) \), mask distribution \( q(\mathbf{M}) \), number of ports \(K\)
\Ensure Trained marginal flows \( \{F_{i,\phi}\}_{i=1}^{6K} \) and copula model \( c_\theta \)

\vspace{1mm}
\State \textbf{Stage 1: Fit marginal flows (unconditional)}
\While{Stage-1 stopping criterion not met}
    \State Sample mini-batch $\{{\bf X}^{(n)}\}_{n=1}^B \sim p_{\text{sim}} \)
    \For{each sample \(n\) and coordinate \(i\)}
        \State Compute \( U_i^{(n)} = F_{i,\phi}\big(X_i^{(n)}\big) \)
        \State Evaluate marginal log-density $\log f_{i,\phi} \big(X_i^{(n)}\big) $
    \EndFor
    \State Compute
    \(
        \mathcal{L}_{\text{marg}}(\phi)
        = \frac{1}{B}\sum_{n=1}^B \sum_i
          \log f_{i,\phi}\big(X_i^{(n)}\big)
    \)
    \State Update \( \phi \) by gradient ascent on \( \mathcal{L}_{\text{marg}}(\phi) \)
\EndWhile

\vspace{1mm}
\State \textbf{Stage 2: Train masked attentional copula}
\While{Stage--2 stopping criterion not met}
    \State Sample mini-batch $\{{\bf X}^{(n)}\}_{n=1}^B\sim p_{\text{sim}}$
    \State Sample masks \( \{\mathbf{M}^{(n)}\}_{n=1}^B \sim q(\mathbf{M}) \)
    \For{each sample \(n\)}
        \State Define
        \( \mathcal{H}^{(n)} = \mathcal{H}(\mathbf{M}^{(n)}),\;
           \mathcal{P}^{(n)} = \mathcal{P}(\mathbf{M}^{(n)}) \)
        \State Compute uniforms \( U_i^{(n)} = F_{i,\phi}\big(X_i^{(n)}\big) \) for all \(i\)
        \State Build encoder inputs from \( \{U_i^{(n)}, X_i^{(n)}, M_i^{(n)}\}_{i\in\mathcal{H}^{(n)}} \)
        \State Build decoder inputs for \( \{i\in\mathcal{P}^{(n)}\} \) (positional/type/mask embeddings)
        \State Run encoder-decoder to obtain conditional densities \( c_{\theta,i}^{(n)}(\cdot) \) for \( i\in\mathcal{P}^{(n)} \)
        \State Accumulate conditional log-likelihood
        $$%\(
            \ell^{(n)}(\theta)
            = \sum_{i\in\mathcal{P}^{(n)}}
              \log c_{\theta,i}^{(n)}
              \big(U_i^{(n)} \mid
                \mathbf{U}_{\mathcal{H}^{(n)}},
                \mathbf{X}_{\mathcal{H}^{(n)}},\mathbf{M}^{(n)}\big)
        $$%\)
    \EndFor
    \State Compute \( \mathcal{L}_{\text{cop}}(\theta) = \frac{1}{B}\sum_{n=1}^B \ell^{(n)}(\theta) \)
    \State Update \( \theta \) by gradient ascent on \( \mathcal{L}_{\text{cop}}(\theta) \)
\EndWhile

%\vspace{1mm}
%\State \textbf{Stage 3: Online fastaFAMA inference with partial observations}
%\State At each symbol time:
%\State \quad Probe a subset of ports \( \mathcal{M} \subset \{1,\ldots,K\} \) and obtain \( \{h_k,r_k\}_{k\in\mathcal{M}} \)
%\State \quad Encode into \( \mathbf{x}_{\mathcal{H}(\mathbf{M})} \), construct mask \( \mathbf{M} \) and uniforms
%      \( U_i = F_{i,\phi}(X_i) \) for \( i\in\mathcal{H}(\mathbf{M}) \)
%\For{\(m = 1,\ldots,M\)}
 %   \State Sample
  %  \( \mathbf{u}_{\mathcal{P}(\mathbf{M})}^{(m)}
  %     \sim c_\theta\big(\cdot \mid
   %       \mathbf{u}_{\mathcal{H}(\mathbf{M})},
    %      \mathbf{x}_{\mathcal{H}(\mathbf{M})},\mathbf{M}\big) \)
    %\State Map back
    %\( \hat{X}_i^{(m)} = %F_{i,\phi}^{-1}\big(U_i^{(m)}\big) \) for all \(i\in\mathcal{P}(\mathbf{M})\)
    %\State Recover complex fields \( \hat{h}_k^{(m)}, \hat{r}_k^{(m)}, \hat{I}_k^{(m)} \) for all \(k\)
%\EndFor
%\State Approximate peraport powers
%\(
 % \widehat{P}_{h,k} =
  %  \frac{1}{M}\sum_m |\hat{h}_k^{(m)}|^2 P_s,
  %\;
  %\widehat{P}_{I,k} =
   % \frac{1}{M}\sum_m |\hat{I}_k^{(m)}|^2
%\)
%\State Compute approximate SINRs
%\(
 % \hat{\gamma}_k
  %= \frac{\widehat{P}_{h,k}}%{\widehat{P}_{I,k} + \sigma_\eta^2}
%\)
%and select
%\(
 % \hat{k}^\star = \arg\max_k %\hat{\gamma}_k
%\)
%or a set of topaSINR ports for combining.
\end{algorithmic}
\end{footnotesize}
\end{algorithm}
%\vspace{-8mm}
%\end{figure}
\end{minipage}

\vspace{-2mm}
\subsection{Complexity and Extensions}\label{subsec:complexity-extensions}
Let $T=3K$ denote the length of the port-major process and $L$ be the number of transformer layers in the copula encoder/decoder, each with $H$ attention heads. For a given mask $\mathbf{M}$ with $|\mathcal{H}(\mathbf{M})| = N_H$ and $|\mathcal{P}(\mathbf{M})| = N_P$, the dominant cost per forward pass is estimated as
\begin{equation}
O\left(L H (N_H^2 + N_P N_H)\right),
\end{equation} which is linear in the number of ports for fixed $M=|\mathcal{M}|$ since $N_H=O(M)$ while $N_P=O(K)$. Thus, even with hundreds of ports, the per-symbol inference complexity is dominated by the number of observed ports rather than by the total FAS size, making the method suitable for large 1D and 2D FAS.

The framework is agnostic to the underlying FAS dimension and channel model. For a 2D planar FAS, one can flatten the $K_x\times K_y$ grid into a single index $k$ via raster scanning, and augment the time embeddings with the physical coordinates $\mathbf{p}_k$. Similarly, finite-scattering models with a random number of paths and AoA distributions can be used during simulation to diversify $p_{\text{sim}}(\mathbf{X})$. The same two-stage attentional copula then learns a unified conditional model that can be deployed under a variety of propagation environments.

Overall, the proposed general attentional-copula-based fast FAMA scheme provides a fully probabilistic, data-driven mechanism to reconstruct the interference field and the unobserved channel coefficients from partial FAS observations, enabling near-ideal fast FAMA operation with a small number of RF chains and without explicit instantaneous knowledge of the interference or transmitted symbols.

\vspace{-2mm}
\section{Simulation Results}\label{sec:simulation_results}
We evaluate the proposed attentional-copula-based interference imputation scheme for fast FAMA. Simulation results are conducted under many different settings by varying the number of users, $U$, the number of observed ports, $M$, and the normalized length of FAS, $W$ under 1D and 2D geometries, and rich and finite scattering channels. The average signal-to-noise ratio (SNR) is set to $10~{\rm dB}$. Unless stated otherwise, all results are obtained from Monte Carlo averaging over independently generated channel and interference realizations. Table~\ref{tab:sim_parameters} summarizes the main simulation parameters used for the system and neural network structures.

\begin{table}[t]
\caption{Simulation and Model parameters.}\label{tab:sim_parameters}
\centering
\resizebox{.9\linewidth}{!}{
\begin{tabular}{l||l}
\thickhline
\textbf{Parameter} & \textbf{Value} \\
\thickhline
\hline
\multicolumn{2}{c}{\emph{FAS and channel model}}\\
\hline
FAS type & 1D linear, 2D planar FAS \\
Total ports $K$ & $200$ \\
Normalized FAS length $W$ & $\{5,10,15,20\}$  \\
Number of interferers $U-1$ & $10$--$200$ \\
Modulation & QPSK \\
\hline
\multicolumn{2}{c}{\emph{Partial observations and dataset}}\\
\hline
Observed ports $M$ & $10$--$60$ \\
Observation mask & Equally-spaced/Random \\
Time-series encoding & Length $3K$ \\
Series channels & $2$ (real and imaginary parts) \\
\hline
\multicolumn{2}{c}{\emph{Model architecture}}\\
\hline
Input channels & $2$ \\
Flow / copula series embedding dim. & $16$ / $16$ \\
Input-encoder layers (flow / copula) & $3$ / $3$ \\
Temporal encoder layers (flow / copula) & $6$ / $6$ \\
Attention heads (flow / copula) & $8$ / $8$ \\
Attention dim. / feedforward dim. & $128$ / $256$ \\
Attentional-copula heads / layers & $8$ / $3$ \\
Attentional-copula MLP layers / dim. & $4$ / $64$ \\
DSF-marginal MLP layers / dim. & $4$ / $64$ \\
Flow layers / hidden dim. (DSF) & $4$ / $64$ \\
\hline
\multicolumn{2}{c}{\emph{Training and inference}}\\
\hline
Optimizer (stage~1 / stage~2) & Adam, $\text{lr}=10^{-5}$ / $5\times10^{-6}$ \\
Epochs (stage~1 / stage~2) & $600$ / $300$ \\
Batches per epoch & $1000$ \\
Batch size (stage~1 / stage~2) & $16$ / $8$ \\
Observed-port range during training & $M\in[10,60]$ \\
\thickhline
\end{tabular}}
\end{table}

\begin{figure}[t]
\centering
\includegraphics[width=\columnwidth]{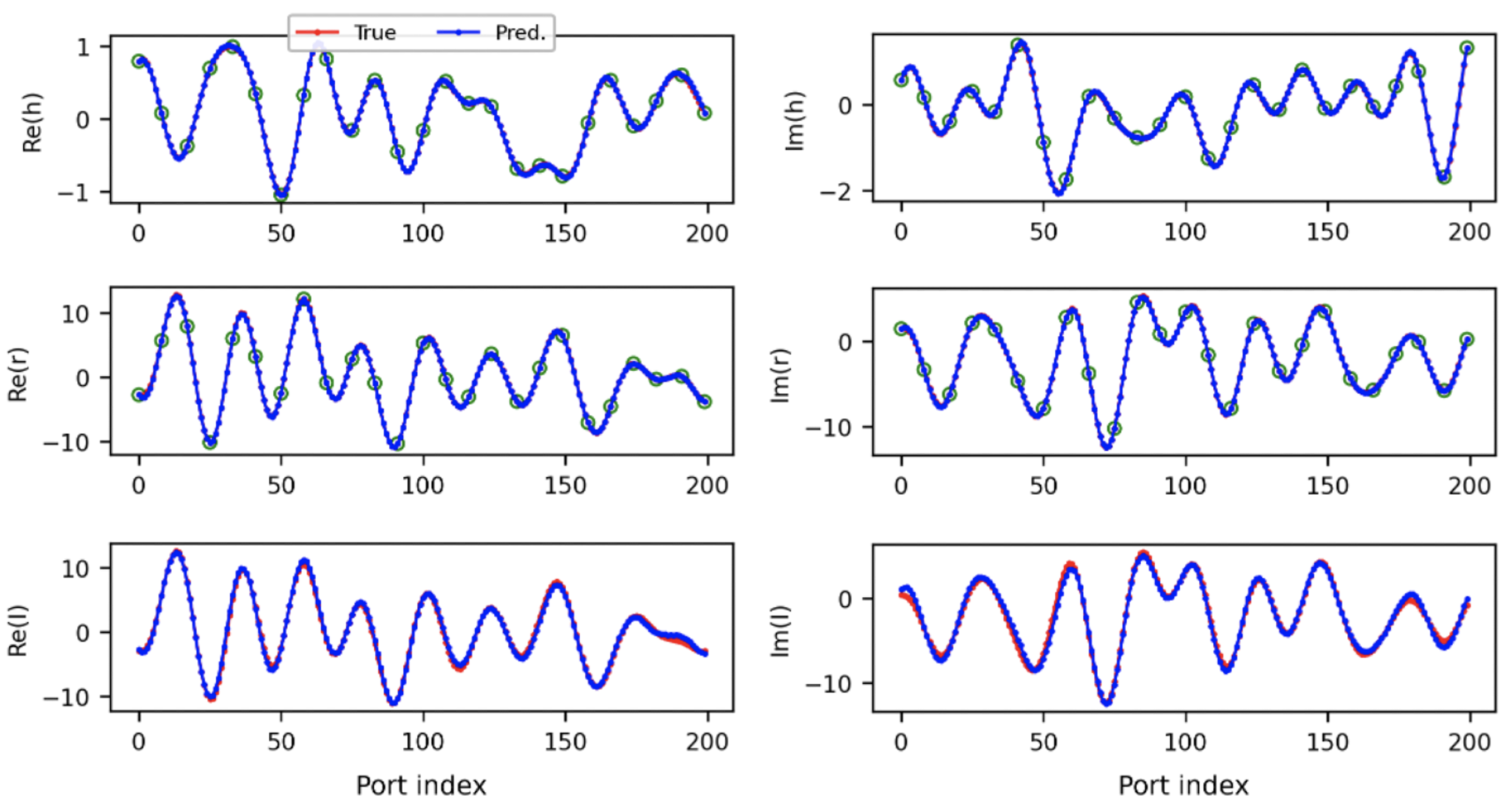}
\caption{Joint prediction of $\Re\{h_k\}$, $\Im\{h_k\}$, $\Re\{r_k\}$, $\Im\{r_k\}$, $\Re\{I_k\}$, and $\Im\{I_k\}$ across the FAS ports for $K=200$ and $M=25$ observed ports. Red curves show the true realization, blue curves the posterior mean prediction, and green circles mark the observed ports.}\label{fig:hrI_pred}
\vspace{-2mm}
\end{figure}

Fig.~\ref{fig:hrI_pred} illustrates the quality of the joint imputation achieved by the proposed scheme for a representative channel snapshot with $K=200$ ports and only $M=25$ observed $(r_k,h_k)$ pairs, assuming $W=10$ and $U=50$ under rich scattering. The six stacked subplots show the real and imaginary parts of $r_k$, $h_k$ and $I_k$, respectively.  Even though only a small fraction of ports are observed, the predicted curves are almost indistinguishable from the true curves over the entire aperture.  The model not only reconstructs the slowly varying desired channel $h_k$ but also accurately tracks the more oscillatory received signal $r_k$ and the heavily interference-dominated field $I_k$.  This qualitative result confirms that the learned joint copula captures the high-dimensional dependence structure among $(r_k,h_k,I_k)$ so well that the unobserved ports can be inferred with almost perfect fidelity from a sparse spatial subset.

\begin{figure}[t]
\centering
\includegraphics[width=.9\columnwidth]{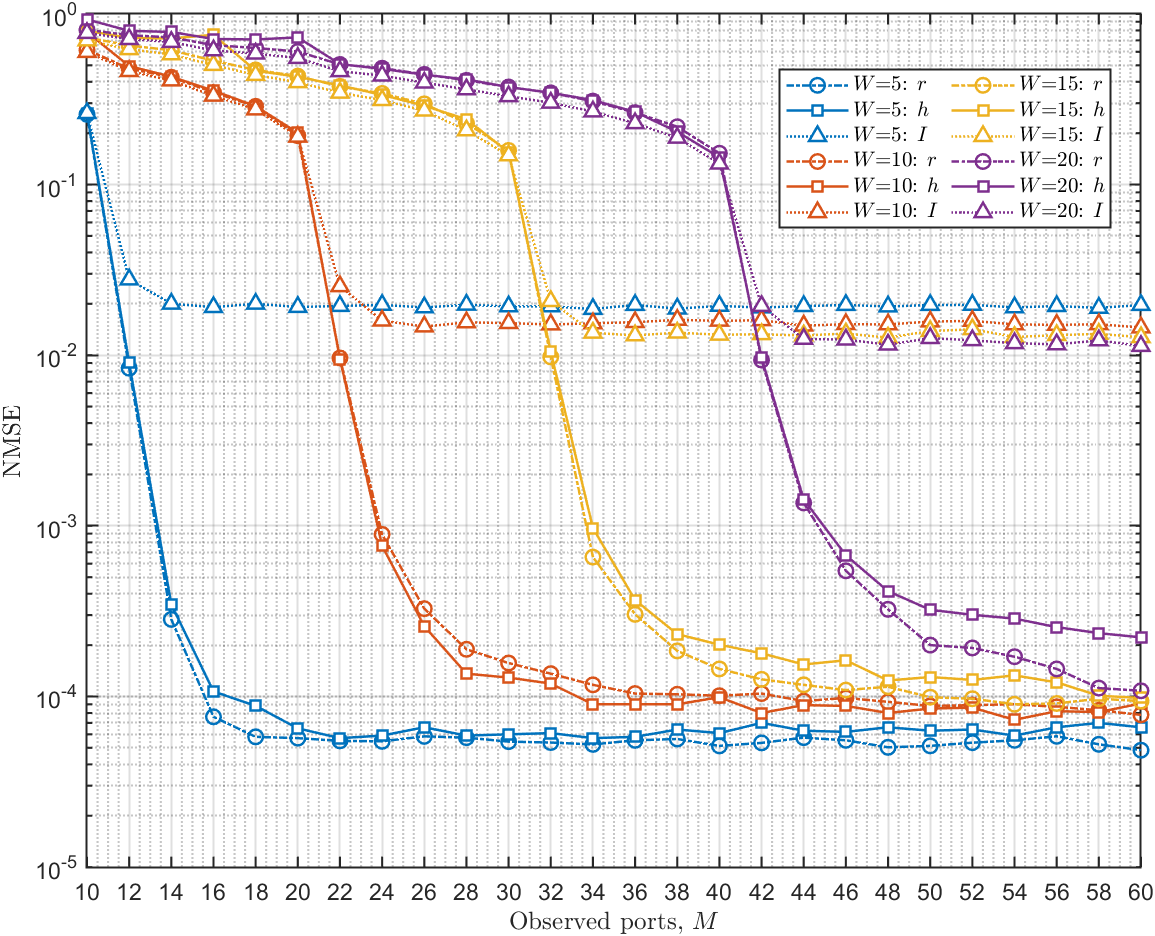}
\caption{NMSE of the imputed $r$, $h$ and $I$ versus the number of observed ports $M$ for different FAS lengths $W$ (1D rich scattering, $K=200$, $U=50$).}\label{fig:nmse_M}
\vspace{-4mm}
\end{figure}

Fig.~\ref{fig:nmse_M} quantifies the reconstruction accuracy in terms of the normalized mean-square error (NMSE) as the number of observed ports, $M$, increases from $10$ to $60$.  For a fixed $W$, the NMSE of both $r$ and $h$ drops rapidly and eventually saturates at a floor on the order of $10^{-4}$ once enough ports are observed.  The required $M$ depends strongly on $W$: for the shortest aperture $W=5$, near-optimal NMSE is achieved already around $M\approx 15$, while for longer apertures $W=10$, $15$, and $20$, the dramatic drop in NMSE occurs roughly at $M\approx 20$, $30$, and $40$, respectively.  These `jumps' are consistent with spatial sampling arguments. A 1D aperture of normalized length $W$ supports on the order of $2W$ independent spatial degrees of freedom (DoF), so once $M$ slightly exceeds this number, the observed ports provide at least one high-quality sample per coherence region along the aperture.  Beyond that point the copula model can reliably interpolate the remaining ports, leading to the sharp transition from poor to near-perfect reconstruction. In contrast, the NMSE of the interference field $I$ varies much more slowly with $M$ and exhibits no pronounced jumps.  Since $I_k$ is the sum of many independent user channels, its instantaneous value is only weakly constrained by a small number of $(r_k,h_k)$ observations; thus the achievable NMSE for $I$ is fundamentally limited even when $M$ is large.

\begin{figure}[t]
\centering
\includegraphics[width=.9\columnwidth]{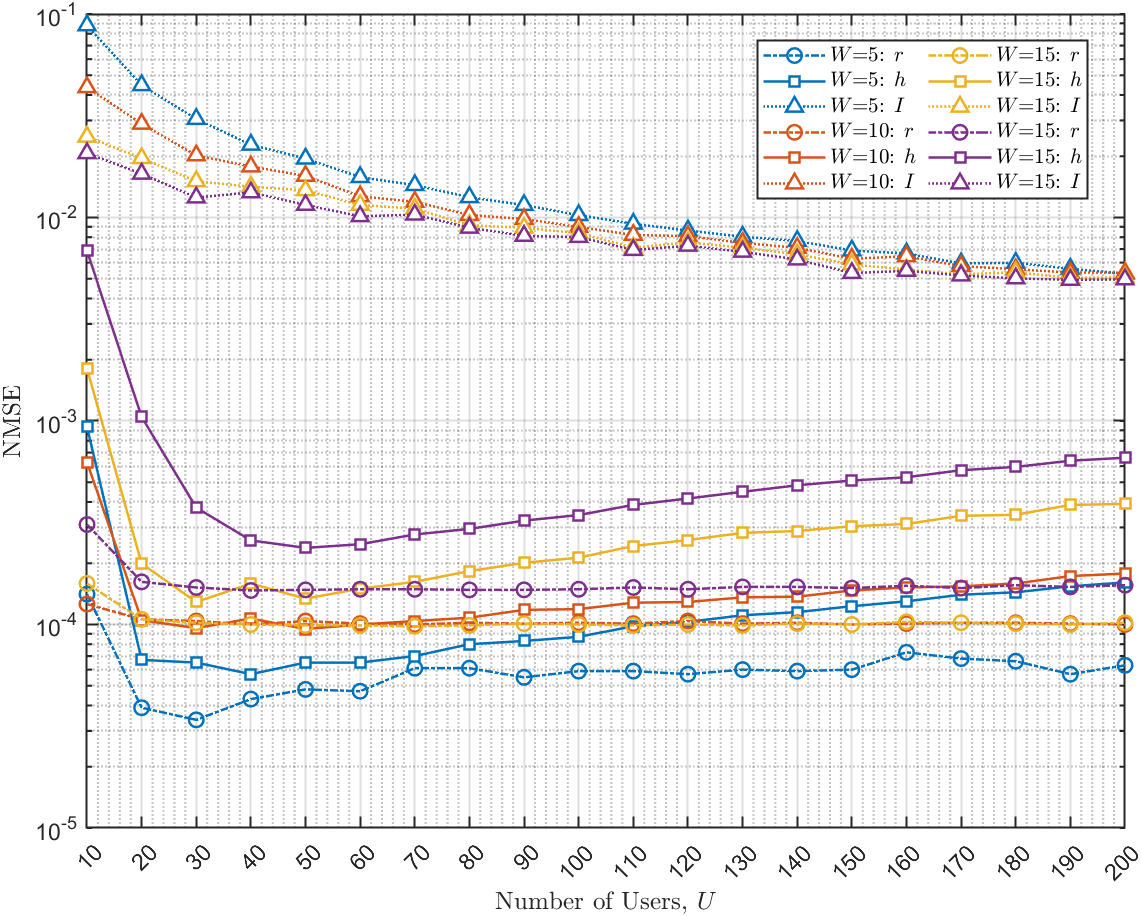}
\caption{NMSE of the imputed $r$, $h$ and $I$ versus the number of interferers $U$ for different FAS lengths $W$ ($K=200$).}\label{fig:nmse_U}
\vspace{-4mm}
\end{figure}

In Fig.~\ref{fig:nmse_U}, we show how the NMSE behaves as the number of interferers $U-1$ ranges from $10$ to $200$ while keeping $K=200$.  For all values of $W$, the NMSE for $r$ and $h$ initially decreases as $U$ increases from $10$ to about $30$, because the aggregate interference $I_k$ becomes closer to a Gaussian random field due to the central limit effect, which makes the joint statistics of $(r,h,I)$ easier to model.  Beyond this point, the NMSE curves either flatten out or increase mildly with $U$ as the interference power grows and the received signal becomes more dominated by strong outliers that are harder to predict exactly.  Nevertheless, the NMSE for both $r$ and $h$ stays well below $10^{-3}$ across the entire range of $U$ once $W\geq 10$, illustrating that the proposed model can robustly infer the desired signal even in heavily interference-limited regimes. Also, the NMSE for $I$ decreases slightly with $U$ since the interference field approaches a more Gaussian and hence easier-to-learn distribution as more users are added.

%To translate instantaneous SINR values into an information-theoretic performance metric that is consistent with the use of a fixed QPSK constellation, we adopt a binary symmetric channel (BSC) sum-rate measure rather than the classical Shannon capacity $\log_2(1+\gamma)$.  The latter assumes Gaussian codebooks, continuous inputs and arbitrarily long blocklengths, which can significantly overestimate the practically achievable rate at the symbol level.

Given the instantaneous SINR $\bar{\gamma}_k$ at the selected port of a UT, the bit error rate for Gray-mapped quadrature phase shift keying (QPSK) with coherent detection is given by
\begin{equation}
p_b(\mathbb{E}[\bar{\gamma}_k])=Q\left(\sqrt{\mathbb{E}[\bar{\gamma}_k]}\right)= \frac{1}{2}\operatorname{erfc}\left(\frac{\sqrt{\mathbb{E}[\bar{\gamma}_k]}}{\sqrt{2}}\right),
\end{equation}
where $Q(\cdot)$ is the Gaussian $Q$-function and $\operatorname{erfc}(\cdot)$ is the complementary error function. Treating each in-phase/quadrature bit stream as a binary symmetric channel (BSC) with crossover probability $p_b$, the corresponding binary entropy is
\begin{equation}
H_b(p_b) = p_b\log_2\frac{1}{p_b} + (1-p_b)\log_2\frac{1}{1-p_b},
\end{equation}
and the per-symbol BSC information rate for QPSK is
\begin{equation}
R_{\rm BSC}(\mathbb{E}[\bar{\gamma}_k])=2\left[1 - H_b\left(p_b(\mathbb{E}[\bar{\gamma}_k])\right)\right]~\mbox{[bits/symbol]}.
\end{equation}
%which captures the loss due to both finite constellation and finite blocklength while remaining a simple function of $\gamma_k$.  

\begin{figure}[t]
\centering
\includegraphics[width=.9\columnwidth]{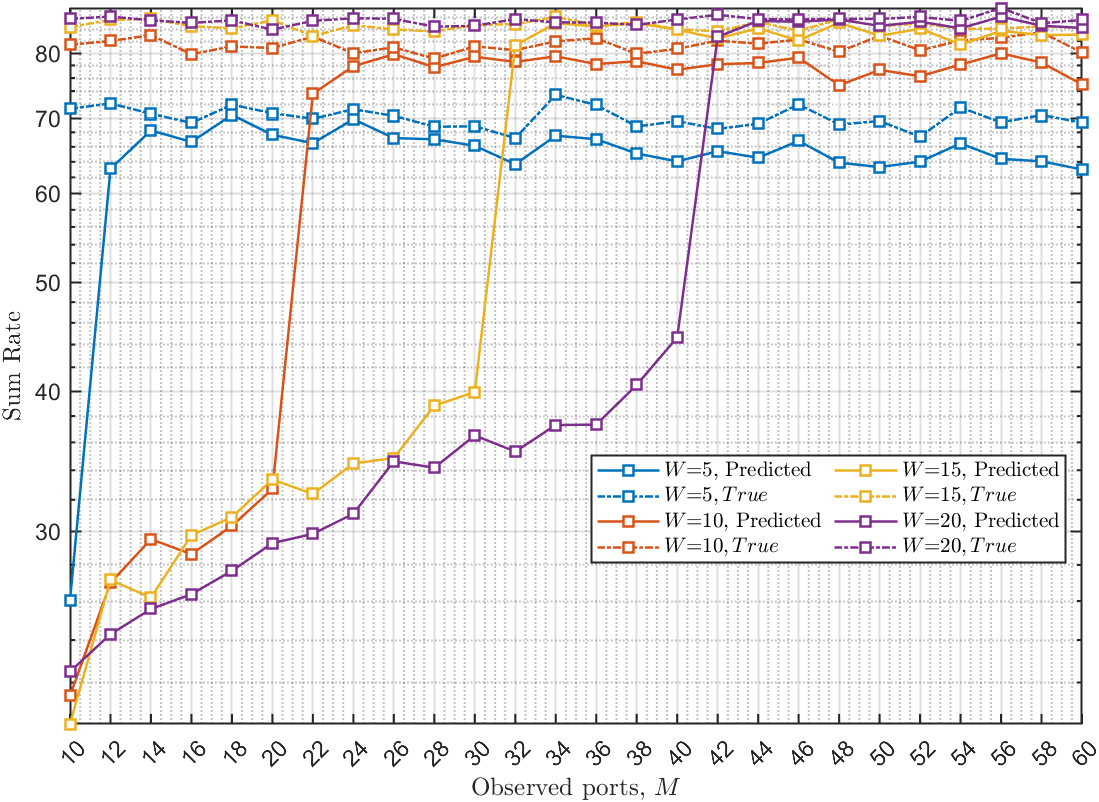}
\caption{BSC sum-rate versus number of observed ports $M$ for different FAS lengths $W$ ($K=200$, $U=50$). `True' curves correspond to oracle selection with perfect knowledge of instantaneous SINR, while `Predicted' curves use the proposed solution-based imputed interference for port selection.}\label{fig:rate_M}
\vspace{-4mm}
\end{figure}

Fig.~\ref{fig:rate_M} reports the BSC sum-rate as a function of the number of observed ports $M$ for $K=200$ and $U=50$. For each FAS length $W$, two curves are shown: a `True' baseline that selects the best port using the exact instantaneous SINRs on all $K$ ports, and a `Predicted' curve where the same selection is performed using the SINRs computed with the imputed interference powers from the proposed solution.  The sum-rate increases monotonically with $M$ and eventually saturates as additional observations provide diminishing returns.  For the shortest aperture $W=5$, the predicted and true curves nearly coincide for all $M$, indicating that even a small number of equally spaced observations suffices to recover the SINR ordering across ports.  For longer apertures ($W=10,15,20$) the predicted curves lie significantly below the oracle when $M$ is small, but the gap closes abruptly once $M$ reaches roughly $M\approx 2W$, precisely where the NMSE curves in Fig.~\ref{fig:nmse_M} exhibited their sharp transitions.  Beyond this point, the proposed solution essentially reproduces the oracle sum-rate, confirming that accurate joint imputation of $(r,h,I)$ directly translates into near-optimal multiuser throughput.

\begin{figure}[t]
\centering
\includegraphics[width=.9\columnwidth]{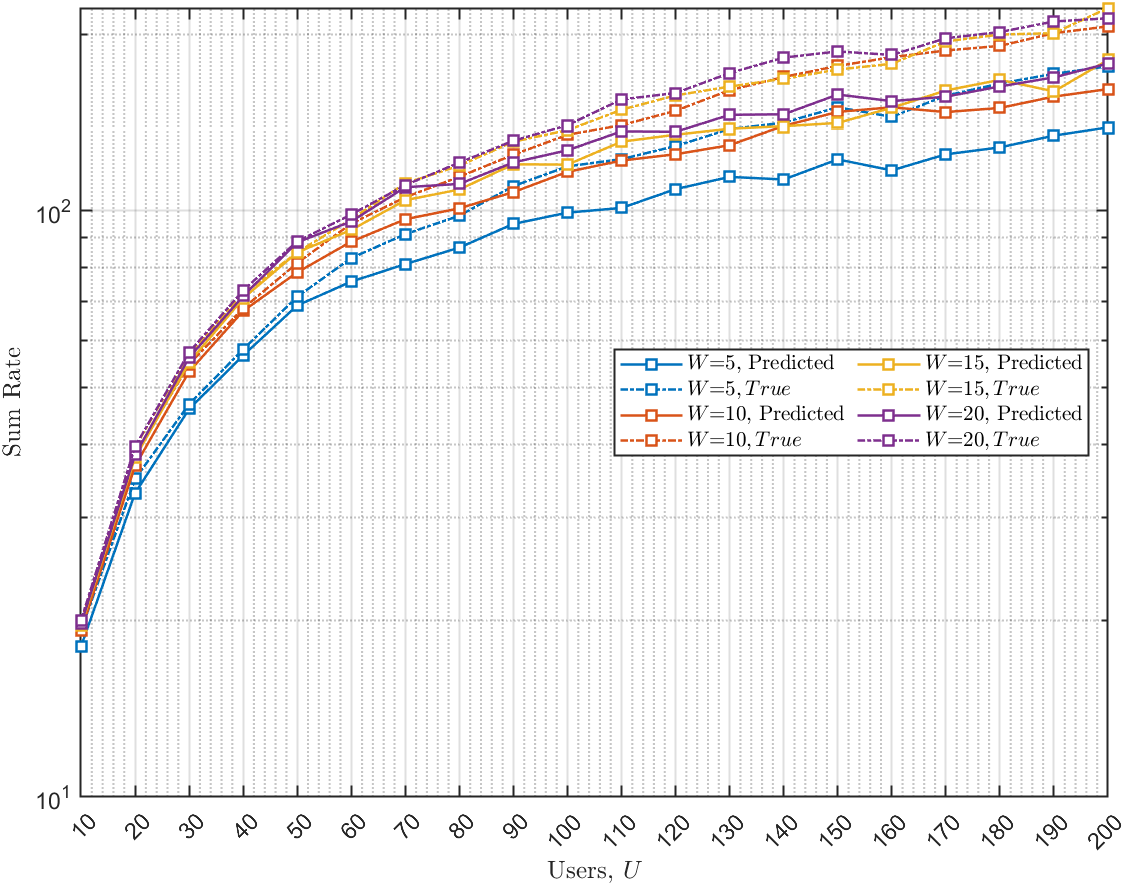}
\caption{BSC sum-rate versus the number of users $U$ for different FAS lengths $W$ ($K=200$).}\label{fig:rate_U}\vspace{-4mm}
\end{figure}

Fig.~\ref{fig:rate_U} depicts the BSC sum-rate as a function of the number of interferers $U$ for $K=200$ and $M=25$.  For each FAS length $W$, the sum-rate grows roughly logarithmically with $U$, adding users both increases the aggregate interference and provides more opportunities to exploit spatial selectivity, but the dominant effect is the linear growth in the number of data streams.  As expected, larger apertures yield significantly higher sum-rates because they offer more spatial DoF, for example, the $W=20$ curves lie well above those for $W=5$ across the entire range of $U$.  Most importantly, the predicted curves track their oracle counterparts extremely closely for all values of $U$ and $W$, with only a small gap at very low $U$.  This demonstrates that once the model has access to a moderate number of observed ports, the proposed interference imputation is robust enough to support accurate port selection even in highly loaded interference-limited scenarios. Note also that the BSC sum-rates for large $W$ do not look very different since the performance is dictated by $K$ which is fixed here.

\begin{figure}[t]
\centering
\includegraphics[width=.9\columnwidth]{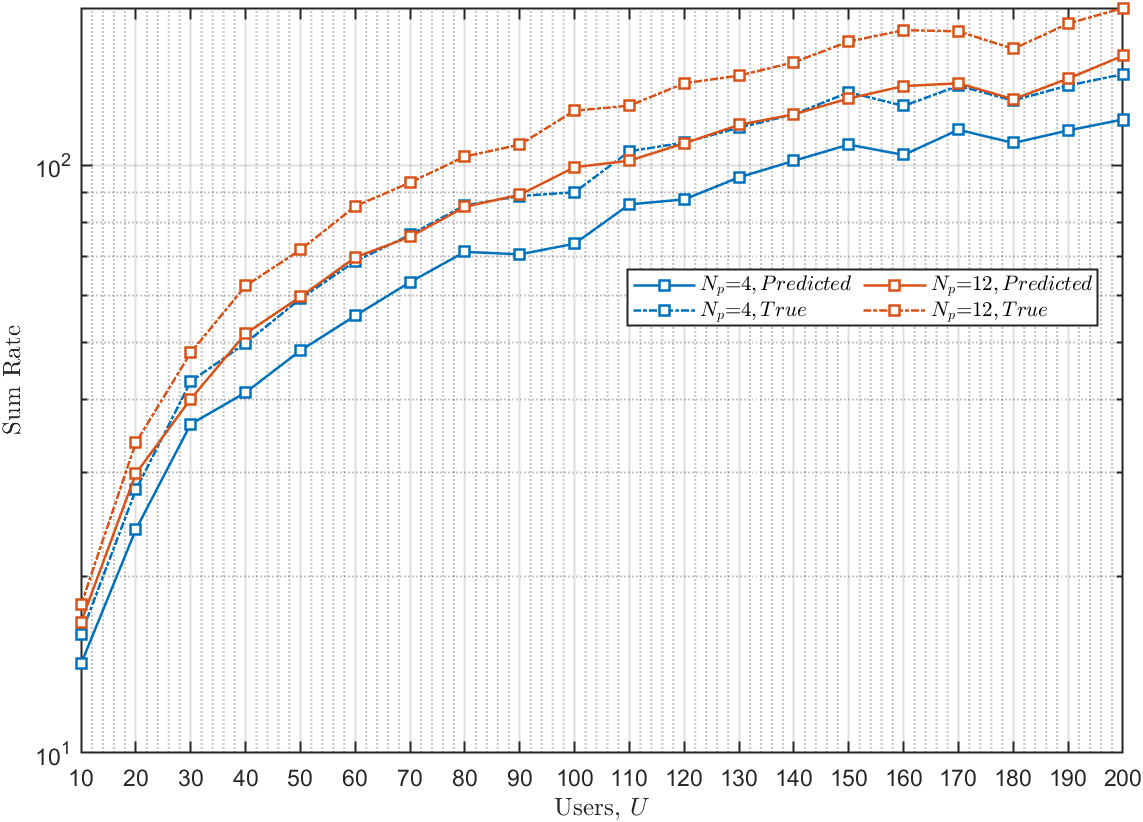}
\caption{BSC sum-rate versus the number of users $U$ for a 1D finite-scattering FAS ($K=200$, $W=10$).}\label{fig:FAMA_RATE_U_finite}
\vspace{-4mm}
\end{figure}

Fig.~\ref{fig:FAMA_RATE_U_finite} depicts the BSC sum-rate versus the number of UTs $U$ for a 1D FAS ($K=200$) with aperture $W=10$ under a finite-scattering model, comparing channels with $N_p=4$ and $N_p=12$ dominant paths. For both values of $N_p$, the sum-rate increases steadily with $U$, as adding users not only raises the interference level but also injects more data streams and creates additional opportunities to exploit spatial selectivity across the aperture. The richer scattering case $N_p=12$ consistently outperforms the more rank-deficient case $N_p=4$, reflecting the larger number of effective spatial DoF provided by additional paths. Across the entire user range, the curves based on the imputed interference (`Predicted') closely follow their oracle counterparts that rely on the true instantaneous SINR (`True'), with only a mild discrepancy at very low and very high $U$. This demonstrates that the proposed copula-aided interference reconstruction remains reliable even in structured, low-rank scattering environments, enabling near-optimal fast FAMA port selection and multiuser throughput.

\begin{figure*}[t]
\centering
\includegraphics[width=0.85\textwidth]{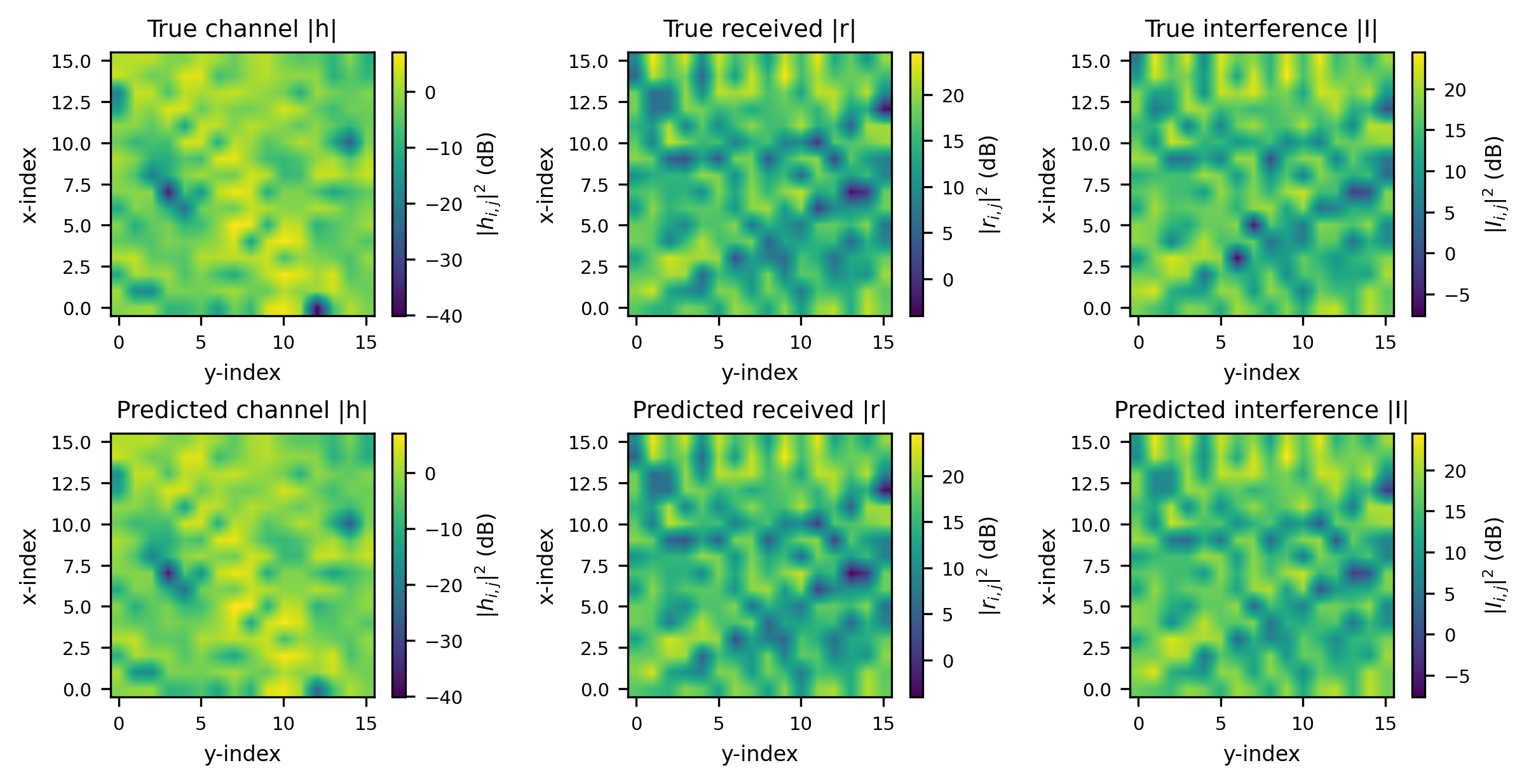}
\caption{True and predicted power maps of the desired channel $|h|$, received signal $|r|$, and aggregate interference $|I|$ over a $16\times 16$ planar FAS ($K=256$ ports) with aperture $W=4\times 4$ wavelengths under rich scattering.}\label{fig:heatmap_2D_hrI}
\vspace{-4mm}
\end{figure*}

Fig.~\ref{fig:heatmap_2D_hrI} illustrates the 2D reconstruction performance of the proposed joint model on a $16\times 16$ planar FAS ($K=256$) with aperture $W=4\times 4$ wavelengths under rich scattering and only $M=40$ observed $(r_{i,j},h_{i,j})$ pairs. For each quantity, the top row shows the true power maps $|r_{i,j}|^2$, $|h_{i,j}|^2$, and $|I_{i,j}|^2$ in dB, while the bottom row displays the corresponding posterior-mean predictions obtained from the proposed model. The predicted heatmaps closely follow their true counterparts, the dominant hot-spots, nulls, and smooth spatial variations of the desired channel $|h|$ are accurately reproduced, and even the more irregular interference-dominated patterns in $|r|$ and $|I|$ are captured with high fidelity. The fact that these fine-grained 2D structures can be recovered from measurements on only a small subset of ports demonstrates that the learned copula effectively exploits the underlying spatial correlation of the rich-scattering field, enabling reliable interpolation of unobserved ports across both dimensions of the aperture.

\begin{figure}[t]
\centering
\includegraphics[width=.9\columnwidth]{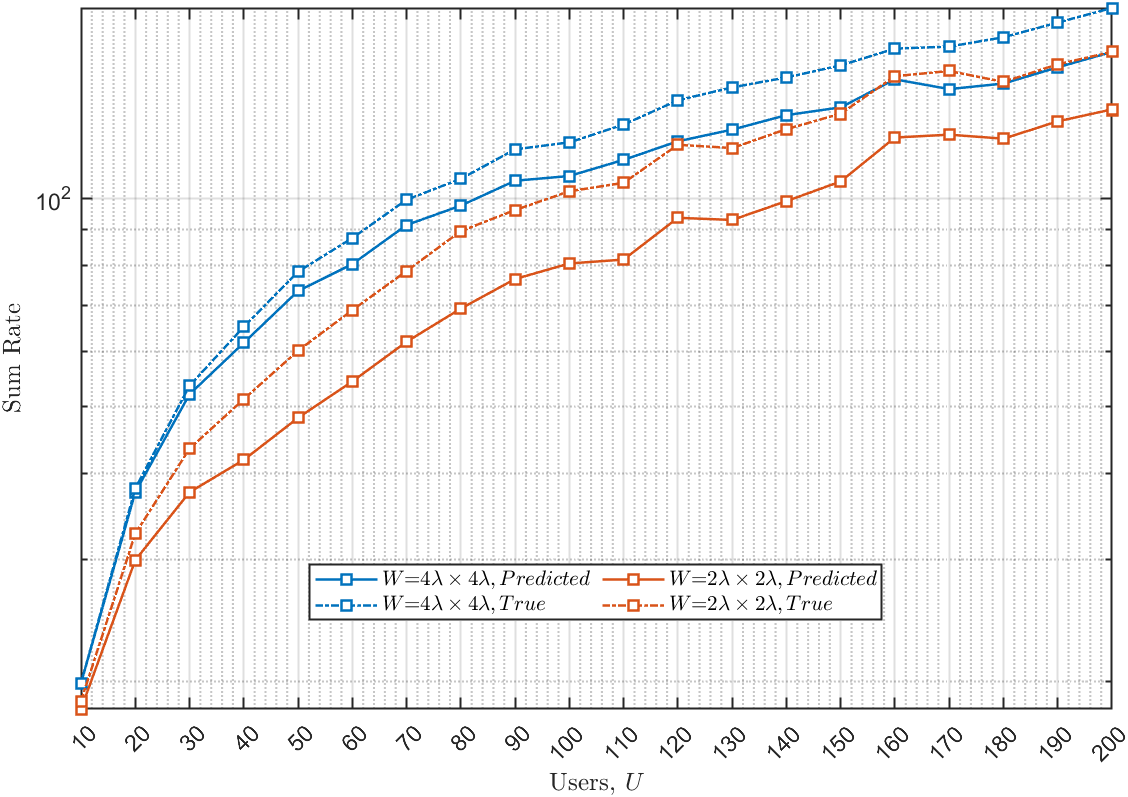}
\caption{BSC sum-rate versus the number of users $U$ for 2D FAS apertures ($K=256$).}\label{fig:rate_U_2D}
\vspace{-4mm}
\end{figure}

Finally, Fig.~\ref{fig:rate_U_2D} shows the BSC sum-rate as a function of the number of UTs $U$ for a $16\times16$ planar FAS ($K=256$) under 2D rich scattering, comparing apertures of size $W=2\lambda\times2\lambda$ and $W=4\lambda\times4\lambda$. For both apertures, the sum-rate grows almost linearly with $U$. Adding UTs increases the aggregate interference but more importantly, provides more data streams and more opportunities for spatial selectivity across the ports. The larger $4\lambda\times4\lambda$ aperture consistently achieves a higher sum-rate than the $2\lambda\times2\lambda$ case, and the gap widens with $U$, reflecting the additional spatial DoF available in the larger 2D FAS. Across the entire user range, the curves based on the imputed interference (`Predicted') closely track their oracle counterparts that rely on the true SINR (`True'). This confirms that even in dense 2D multiuser settings, the proposed copula-aided interference reconstruction is accurate enough to support near-optimal port selection and multiuser throughput.

\vspace{-2mm}
\section{Practical Implication}\label{sec:practice}
The numerical results above in Section \ref{sec:simulation_results} shed light on how to improve existing MIMO technologies and offer insights into the role of FAS. In particular, existing MIMO systems are based on FPAs with independent signals for maximum diversity. Nonetheless, our results suggest something different. It is actually possible to have $M\approx 2W$ FPAs to take signal samples $\{r\}$ and then apply the proposed general attentional copula method to infer $K\gg M$ signal samples over the given aperture. In so doing, the principle of fast FAMA can be used without necessarily using FAS at the UT, to mitigate the severe interference, still without any of precoding and SIC. That said, $M$ can still be considered too large for UT as the cost of an RF chain is high. In this case, FAS is the unique technology that can allow the UT to take multiple signal samples on much less or even one RF chain. Additionally, there may be benefits of unevenly sampling the signals over the aperture using FAS as opposed to multiple FPAs, which deserves investigation.

\vspace{-2mm}
\section{Conclusion}\label{sec:conclusion}
This paper addressed a key practicality gap in fast FAMA systems, the assumption that each UT can measure the instantaneous SINR on all FAS ports at every symbol instance. Specifically, we recast fast FAMA as a joint imputation problem over the complex triplets $(r_k,h_k,I_k)$ and developed a \emph{copula-aided FAMA} framework that reconstructs the interference field and unobserved channel coefficients from partial observations. The proposed scheme employs a two-stage attentional copula time-series model, built on marginal normalizing flows and a transformer-based copula. During training, random observation masks emulate realistic sensing of only $M$ ports per symbol, enabling the model to learn conditional laws that generalize to practical hardware constraints. Our simulation results for 1D/2D FAS geometries, under both rich and finite scattering, showed that the learned model captures the spatial structure of the CSI and interference fields with high fidelity, realizing near-genie fast FAMA under practical settings.

\ifCLASSOPTIONcaptionsoff
  \newpage
\fi

%\vspace{-2mm}
\bibliographystyle{IEEEtran}
%\bibliography{bibliography2.bib}
% Generated by IEEEtran.bst, version: 1.14 (2015/08/26)
\section*{REFERENCES}
\def\refname{\vadjust{\vspace*{-1em}}}

\end{document}